\documentclass[epsfig, usenatbib]{mn2e}
\usepackage{epsfig}
\usepackage{natbib}
\usepackage{graphicx}
\usepackage{fixltx2e}
\usepackage{multirow}
\usepackage{amsmath}
\usepackage{amssymb}
\usepackage{url}
\title{The different star-formation histories of blue and red spiral and elliptical galaxies}

\author[Tojeiro et al.]{
 \parbox{\textwidth}{Rita Tojeiro$^1$\thanks{E-mail: rita.tojeiro@port.ac.uk},  
 Karen L. Masters$^1$,
 Joshua Richards$^2$,
 Will J. Percival$^1$,\\
 Steven P. Bamford$^3$,
 Claudia Maraston$^1$,
 Robert C. Nichol$^1$,
 Ramin Skibba$^4$,\\
 Daniel Thomas$^1$
 }\vspace{3mm}\\
 $^1$Institute of Cosmology and Gravitation, Dennis Sciama Building, University of Portsmouth, Burnaby Road, Portsmouth, PO1 3FX, UK \\
$^2$Department of Physics, Imperial College London, London, SW7 2AZ, UK \\
$^3$School of Physics \& Astronomy, University of Nottingham, University Park, Nottingham NG7 2RD \\
$^4$Center for Astrophysics and Space Sciences, Department of Physics, University of California, 9500 Gilman Dr., La Jolla, San Diego,\\
		CA 92093, USA
}

\def\gs{\mathrel{\raise1.16pt\hbox{$>$}\kern-7.0pt %
\lower3.06pt\hbox{{$\scriptstyle \sim$}}}}         %
\def\ls{\mathrel{\raise1.16pt\hbox{$<$}\kern-7.0pt %
\lower3.06pt\hbox{{$\scriptstyle \sim$}}}}         %

\begin{document}

\maketitle

\begin{abstract}
{We study the spectral properties of intermediate mass galaxies ($M_* \sim 10^{10.7} M_\odot$) as a function of colour and morphology. We use Galaxy Zoo to define three morphological classes of galaxies, namely early-types (ellipticals), late-type (disk-dominated) face-on spirals and early-type (bulge-dominated) face-on spirals. We classify these galaxies as blue or red according to their SDSS $g-r$ colour and use the spectral fitting code VESPA to calculate time-resolved star-formation histories, metallicity and total starlight dust extinction from their SDSS fibre spectra. 

We find that red late-type spirals show less star-formation in the last 500 Myr than blue late-type spirals by up to a factor of three, but share similar star-formation histories at earlier times. This decline in recent star-formation explains their redder colour: their chemical and dust content are the same. We postulate that red late-type spirals are recent descendants of blue late-type spirals, with their star-formation curtailed in the last 500 Myrs. The red late-type spirals are however still forming stars $\simeq17$ times faster than red ellipticals over the same period.

Red early-type spirals lie between red late-type spirals and red ellipticals in terms of recent-to-intermediate star-formation and dust content. Therefore, it is plausible that these galaxies represent an evolutionary link between these two populations. They are more likely to evolve directly into red ellipticals than red late-type spirals, which show star-formation histories and dust content closer to blue late-type spirals.

Blue ellipticals show similar star-formation histories as blue spirals (regardless of type), except they have formed less stars in the last 100 Myrs. However, blue ellipticals have different dust content, which peaks at lower extinction values than all spiral galaxies. Therefore, many blue ellipticals are unlikely to be descendants of blue spirals, suggesting there may not be single evolutionary path for this group of galaxies.

}
\end{abstract}


\title{Colour and morphology}

\section{Introduction}  \label{sec:intro}
It has been know for almost a century (Hubble 1922) that galaxies come in two basic types: spiral galaxies which exhibit a disc of stars, usually with spirals arms; and elliptical galaxies which are predominantly spheroidal. Furthermore the spiral galaxies form a sequence, largely defined by growing bulge size (Sc-Sb-Sa) culminating in S0 (or lenticular galaxies) with extremely large bulges, which are often grouped with ellipticals (and collectively called early-types)\footnote{Throughout this paper we use 'elliptical galaxy' and 'early-type galaxy' interchangeably, noting that this class contains most S0s.}. This morphological bimodality in the galaxy population has been the basic observation to be explained by studies of galaxy evolution since it has been understood that the objects were galaxies external to our own. 

While it had long been noted that ellipticals tended to be redder than spirals (e.g. \citealt{Hubble26,Holmberg58}), only quite recently it was noticed that the local galaxy population demonstrates a strong bi-modality in optical colour (e.g. \citealt{StratevaEtAl01, BlantonEtAl03, BaldryEtAl06}) with a ``blue cloud" and ``red sequence", and an underpopulated ``green valley" in between them. This bimodality correlates extremely well with the type, or morphology of galaxies, with late types (spirals) predominantly found to be in the blue cloud, and early types on the red sequence (e.g. \citealt{StratevaEtAl01, Conselice06, MignoliEtAl09}). Because of this it is now common practice to divide the galaxy population using colour on a colour-magnitude diagram (CMD). 

Observations at high redshift show that the red sequence builds up over time (e.g. \cite{BundyEtAl06}), so one of the main tasks of galaxy evolution studies has become to develop an understanding of the mechanisms by which galaxies progress from the blue cloud to the red sequence (e.g. \citealt{BellEtAl07, SkeltonEtAl12}). The tight relationship between colour and morphology and other supporting evidence led to popularity for the merger origin of red elliptical galaxies; however problems with this picture are emerging (e.g. \citealt{OeschEtAl10, CisternasEtAl11}). 


It is, of course, possible to find samples of galaxies that are the exception to the tight relationship between colour and morphology. Late-type or disc galaxies with predominantly red colours and early-type galaxies with unusually blue colours have been reported since the 70s (e.g. the work of van den Bergh (1976) on anemic spirals in the Virgo cluster, and \citealt{Huchra77} on blue Markarian galaxies). The sizes of these samples were small - and even if detailed number densities were difficult to obtain due to completeness issues, it was clear these systems are rare in the local Universe. This means of course that the processes that lead to these unusual objects must either be rare, or produce these unusually coloured galaxies for only short times. In either case, both red spirals and blue ellipticals stand out as great laboratories for galaxy evolution experiments aimed at increasing our understanding of the formation of early-type galaxies and the build up of the red sequence. 

With the advent of large imaging and spectroscopic surveys, studies of the galaxy population entered a new era of statistical power. In particular, the Sloan Digital Sky Survey (SDSS) imaged over one quarter of the sky, and the SDSS-I and SDSS-II projects obtained spectra and redshifts for over one million galaxies in the nearby Universe \citep{AbazajianEtAl09}. However, due to the need for visual inspection to identify morphology studies of the correlation between colour and morphology remained limited in size \citep{FukugitaEtAl07}, or had to rely on automatically measured proxies for morphology (e.g. \citealt{Conselice06}). 

Galaxy Zoo (GZ; \citealt{LintottEtAl08, LintottEtAl11}) extended the statistical power of SDSS to morphology studies by asking members of the public to visually classify the 1 million galaxies found in the SDSS Main Galaxy Sample (MGS; \citealt{StraussEtAl02}). GZ was motivated in part by a desire to identify a larger sample of blue ellipticals than had been found by the MOSES project (\citealt{SchawinskiEtAl07a}a,  Schawinski et al. 2007b, Schawinski et al. 2009a, \citealt{ThomasEtAl10}), and indeed one of the first results showed that about 6\% of the low redshift early-type galaxy population are blue (\citealt{SchawinskiEtAl09b}b), with the fraction of blue ellipticals increasing to as much as 12\% towards lower mass galaxies in lower density regions \citep{BamfordEtAl09}. This work also demonstrated the difficulty of identifying blue early types with automated structural parameter measurements. 

 The GZ data also revealed a significant population of red spirals \citep{BamfordEtAl09, SkibbaEtAl09c}. With a simple selection of all spiral/disc galaxies, these works suggested around 20\% of these galaxies are to be found in the red sequence (with an increasing fraction in intermediate density regions). The fraction of red spirals was seen to increase with environment, even at fixed stellar mass, suggesting some environmental process is surpressing star-formation in spiral galaxies, beyond a simple change in the mass function with environment \citep{BamfordEtAl09}. At around the same time \cite{WolfEtAl09} identified a sample of red spirals in the Space Telescope A9012/Galaxy Evolution Survey (STAGES), using HST morphologies. They demonstrated that these optically red spirals had a non-zero (but significantly lowered) star formation rate when compared to blue spirals. All three works included highly inclined spirals in their samples. 
 
A substantial number of normally star forming spirals can found in the red sequence due to the effects of inclination dependent dust reddening \citep{MallerEtAl09, MastersEtAl10a}.  It has also long been known that spiral galaxies with large bulges can be quite red (e.g. Haynes \& Roberts 1994, James et al. 2008) and \cite{MastersEtAl10a} demonstrated that the trend in colour due to increasing bulge size is comparable in size to inclination dependent reddening. For these reasons \cite{MastersEtAl10} selected a new sample of face-on spirals from GZ with small bulges (identified via light profile fits) to study the properties of spirals with instrinsically red discs. Of these face-on late-type spirals objects they found about 6\% lay in the red sequence (increasing to 30\% in more bulge dominated spirals). This work showed that red spirals were not more dust reddened than similar blue spirals, and while they had less star formation than blue spirals they were not completely passive objects (which was confirmed by \cite{Cortese12} looking at the UV properties of the sample). The fraction of late-type spirals in the red sequence was observed to increase substantially with stellar mass, but at all stellar masses the red spirals were observed to have older stellar populations and less recent star formation than the carefully selected late-type blue spirals comparison sample. 
 
 While red spirals are relatively rare in the local universe they may form an important evolutionary pathway between the blue cloud and red sequence. \cite{BundyEtAl10} used data from the COSMOS survey to study red sequence galaxies with disc-like morphology at high redshift, and concluded from the relative fractions over cosmic history, that a significant fraction of discs moving from the blue cloud to the red sequence (60\%) may pass through a red spiral phase. \cite{RobainaEtAl12} compared the stellar populations of a samples of galaxies with either spirals or elliptical morphology (from GZ) which were selected to be quiescent by requiring no H$\alpha$ emission in their SDSS fibre spectra and a colour-colour selection designed to remove dust reddened objects (e.g. Williams et al. 2009; \citealt{HoldenEtAl12}). These red spirals by definition are more passive and quiescent than the sample of \cite{MastersEtAl10} and, in addition, no selection on spiral bulge size was made. The work of \cite{RobainaEtAl12} found (in agreement with the work of \citealt{ThomasEtAl06}) that the stellar populations in the two samples (in age, metallicity and $\alpha$-enhancement) were statistically indistinguishable.  





In this paper we present the results of a detailed spectral analyses of carefully selected galaxy samples that span the full colour-morphology space, which we define using SDSS photometry and GZ public morphological classifications. We use the public database of \cite{TojeiroEtAl09} to obtained minimally parametric and detailed star-formation histories, metallicity histories and dust content for the galaxies in our sample. The database of \cite{TojeiroEtAl09} is the product of applying the VErsatile SPectral Analyses (VESPA, \citealt{TojeiroEtAl07}) code to SDSS's Data Release 7 (DR7). The VESPA algorithm applies full spectral fitting to find the best combination of stellar populations of different ages and metallicities, modulated by a dust extinction curve, that best fits each individual spectrum. The age resolution of the star-formation histories returned by VESPA can help disentangle the evolutionary paths of the four samples in significantly more detail than previous studies.

This paper is organised as follows: we define and characterise our samples in Section \ref{sec:data}, and we present our methodology in Section~\ref{sec:method}. In Section~\ref{sec:results} we present our results. Finally, in Section~\ref{sec:discussion} we summarise, discuss and conclude. 


\section{Data}\label{sec:data}

The Sloan Digital Sky Survey (SDSS) I/II has imaged over one quarter of the sky using a
dedicated 2.5m telescope in Apache Point, New Mexico \citep{GunnEtAl06}. For details on
the hardware, software and data-reduction see \citet{YorkEtAl00} and
\citet{StoughtonEtAl02}. In summary, the survey was carried out on a
mosaic CCD camera \citep{GunnEtAl98} and an auxiliary 0.5m telescope for photometric
calibration. Photometry was taken in five bands: $u, g, r, i$ and $z$ \citep{FukugitaEtAl96}, and magnitudes corrected for Galactic extinction using the dust maps of \cite{SchlegelEtAl98}. The photometric and spectroscopic scope of SDSS I/II is being increased with SDSS-III \citep{EisensteinEtAl11}. In this paper we use data exclusively from SDSS-I/II, which we will refer to as SDSS for simplicity.

Galaxies were selected for spectroscopic follow-up mainly via two distinct algorithms. The Main Galaxy Sample (MGS) is a highly complete ($>99\%$)  $r-$band selected sample of galaxies with $r_p>17.77$ ($r_p$ is the Petrosian magnitude; \citealt{Petrosian76, StraussEtAl02}) and a mean redshift $\bar{z}\approx 0.1$ (6\% of the galaxies in the sample do not have measured redshifts due to fibre-collisions). The Luminous Red Galaxy sample (LRG) targets mainly massive early-type galaxies with $0.15 < z < 0.5$ and a lower number density. Spectra are obtained via optical fibres with a diameter of 3 arc-seconds.

The Galaxy Zoo project (Lintott et al. 2008)\footnote{www.galaxyzoo.org} made available images of SDSS MGS galaxies via an internet tool and in its first version (GZ1, which ran in 2007/2008)\footnote{The original website is still available, at  zoo1.galaxyzoo.org} asked members of the public to classify the galaxies simply as ``spiral", ``elliptical", "merger" or "star/don't know". In addition it asked them to identify the direction of rotation of the spiral arms (or indicate that they could not tell) for spiral galaxies. A median of 20 people classified each galaxy in GZ1 and these classification probabilities (i.e. $p_{\rm sp}$ representing the weighted fraction of classifiers who called the galaxy a spiral, and $p_{\rm el}$: the corresponding quantity for the elliptical classification, etc.) were made publicly available in Lintott et al. (2011). We impose a volume limit on the sample of $0.03<z<0.085$ with the lower limit set to reduce the impact of peculiar velocities on distance errors, and the upper limit set to limit the redshift dependent GZ classification bias discussed in the Appendix of Bamford et al. (2009). MGS sample galaxies have $r_p<17.77$ so galaxies with $M_r<-20.17$ are observable throughout this volume. In order to reduce the impact of inclination-dependent reddening moving intrinsically blue spirals observed at high inclinations into the red spiral samples, we also impose an axial ratio limit of $\log (a/b) < 0.2$ that removes highly elongated objects (see also Masters et al. 2010a) . Finally, we use only the ``clean" sample of GZ (Lintott et al. 2011) to identify spirals or early-type galaxies, remove any galaxy which does not have either $p_{\rm sp}>0.8$ or $p_{\rm el}>0.8$. Those with $p_{\rm sp}>0.8$ are identified as spirals, and those with $p_{\rm el}>0.8$ are identified as early-types. 

\begin{figure}
\begin{center}
\includegraphics[width=3.5in]{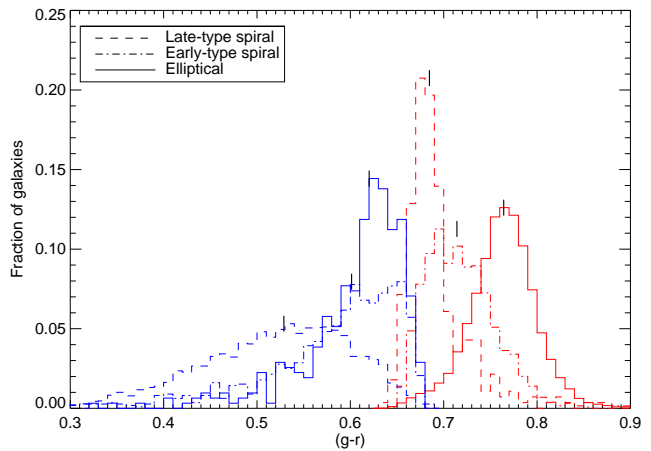}
\caption{The normalised distribution of colours of the six samples: late-type spiral galaxies in dashed lines, early-type galaxies in dot-dashed lines, and elliptical galaxies in solid lines. Red and blue galaxies are shown in red (the rightmost histograms) and blue (the leftmost histograms) respectively. The small black vertical lines show the median of each distribution, as tabulated in Table \ref{tab:colours}. We note how, even though we have only two labels for colour - red and blue - the range in colours within these labels is significant.
In this paper we used detailed spectral analyses to investigate the properties of these four samples in significantly more detail than what is allowed by using colour alone. } 
\label{fig:grplot_bulgy}
\end{center}
\end{figure}

\begin{figure*}
\begin{center}
\includegraphics[width=7in]{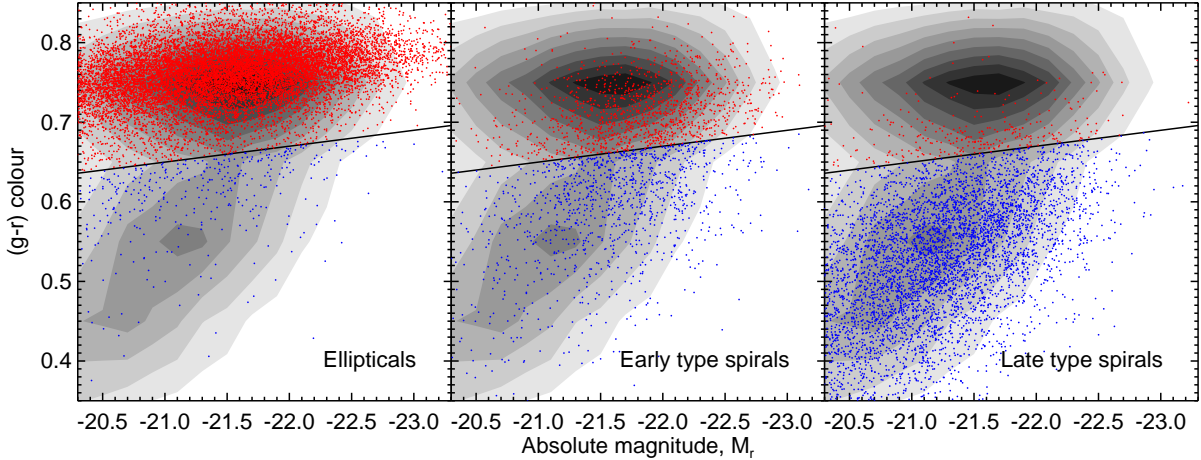}
\vspace{-6cm}
\caption{The colour classification plane for the samples used in this paper. In all three panels the contours are the same, and show the location of all galaxies in the face-on ($\log (a/b)< 0.2$) volume limited ``clean" GZ1 sample described in Masters et al. (2010). The solid line corresponds to the cut found in \protect\cite{MastersEtAl10}, and sits one standard deviation blueward of the fit to the Galaxy Zoo elliptical ($p_{\rm el}>0.8$) colour-magnitude relation. The different panels show the location of different morphological sub-samples from GZ1 classifications in this diagram (specifically: ellipticals are defined as having $p_{\rm el}>0.8$, spirals with $p_{\rm sp}>0.8$ and visible spiral arms, and with early/late type spirals split by fracdev $=0.5$). Points are coloured red where they lie above the colour cut and blue below it.} 
\label{fig:colourmagdiagram2}
\end{center}
\end{figure*}

To define red or blue galaxies we use the fit to the red sequence of Galaxy Zoo ellipticals reported in \cite{MastersEtAl10} which gives $(g-r) = 0.73 - 0.02 (M_r + 20)$, with a 1$\sigma$ scatter of 0.1 mag\footnote{SDSS recommends the use of model magnitudes to obtain the best colours of extended objects. All colours mentioned in this paper are SDSS model magnitude colours.} (colours and magnitudes have been k-corrected to $z=0$). Following \cite{MastersEtAl10} we define a galaxy as ``red" if it has a $(g-r)$ colour redder than $1\sigma$ blueward of the red sequence. Galaxies bluer than this we call ``blue". We show the normalised colour distributions (k-corrected to $z=0$) of all the samples used in this paper in Fig.~\ref{fig:grplot_bulgy} and Table \ref{tab:colours} provides a summary. Fig.~\ref{fig:colourmagdiagram2} shows the colour-magnitude relation and the colour-classification relation for all samples.
  
 The choice of magnitude used to construct the optical colour has an impact on which galaxies are selected as blue or red. We note that a selection based on petrosian magnitude colours, rather than model magnitude colours, results in an overall shift of all galaxies towards the blue as this colour is less dominated by the bright central regions of galaxies which tend to be redder. Galaxies with the largest colour gradients have the bigger difference between model and petrosian magnitude based colours - resulting in a shift of late- and early-type spiral galaxies more to the blue than elliptical galaxies. In our specific case (when the dividing line is defined as 1 sigma bluer than the elliptical red sequence), this reduces the size of the red spiral samples to by approximately half. The effect is the most pronounced for red early-type spirals which have the largest colour gradients. We have rerun all the analysis for this selection and can confirm that all qualitative conclusions based on the model magnitude selection are robust, and changes in the quantitative conclusions are negligible when compared to the uncertainty introduced by SSP model selection, the effects of dust, and fibre aperture corrections.
  
\begin{table}
\begin{tabular}{|c|l|c|c|c|c|}
\hline
 Colour & sample &median & mean & 1$\sigma$ & size \\ \hline
 \multirow{4}{*}{$g-r$} & Red ellipticals & 0.76 & 0.76 & 0.09 & 13959 \\ 
				  & Red early-type spirals & 0.71 & 0.72 & 0.04 & 1265 \\
				 & Red late-type spirals & 0.68 & 0.70 & 0.06 & 294\\
				  & Blue ellipticals & 0.62 &  0.60 & 0.06 & 381 \\
 				  & Blue early-type spirals & 0.60 & 0.58 & 0.08 & 1144 \\ 
				  & Blue late-type spirals & 0.53 & 0.52 & 0.08 & 5139\\ \hline
 \multirow{4}{*}{$u-r$} & Red ellipticals & 2.55 & 2.54 &  0.21&13959\\
 				   & Red early-type spirals & 2.33 & 2.33 & 0.20  & 1265 \\				   
				   & Red late-type spirals & 2.15 & 2.16 & 0.18& 294\\
				   & Blue ellipticals & 1.99 & 1.95 & 0.38&  381\\
 				  & Blue early-type spirals & 1.85 & 1.82 & 0.36 & 1144 \\ 
				   & Blue late-type spirals & 1.71 & 1.71 & 0.25&  5139\\ \hline
 \label{tab:colours}
 \end{tabular}
 \caption{Mean and median colours (k-corrected to $z=0$) of the six galaxy samples described in this paper, and the size of each sample. The full normalised colour distributions can be seen in Fig.~\ref{fig:grplot_bulgy}.}
 \end{table}

 Finally, in the spiral subset, in order to identify true spiral galaxies we require that the spiral arms were clearly visible to GZ volunteers (by imposing $p_{\rm arms}>0.8$). These are then divided into two spiral subsets: early- and late-type spirals, defined as follows. The late-type spirals consist of the ``discy" spiral subset discussed in \cite{MastersEtAl10}, who used information from the light profile shape to remove spiral galaxies with a significant bulge component. This avoids selecting spiral galaxies that are primarily -- or exclusively -- red due to a strong bulge component. Nonetheless, early-type (or bulge-dominated) spirals form a significant part of the spiral-galaxy population, particularly at the higher mass end (e.g.  Haynes \& Roberts 1994, \citealt{BernardiEtAl10}) and it is interesting to compare our sample of late-type spirals to an equivalent sample of bulgy early-type spirals, with a prominent bulge component. Here we define a bulge dominated spiral as a galaxy visually selected as a spiral by Galaxy Zoo, but having a light profile best fit by a de Vaucouleur profile (i.e. {\tt fracdeV} $> 0.5$ from the SDSS pipeline).
  
\begin{figure*}
\begin{center}
\includegraphics[width=6in]{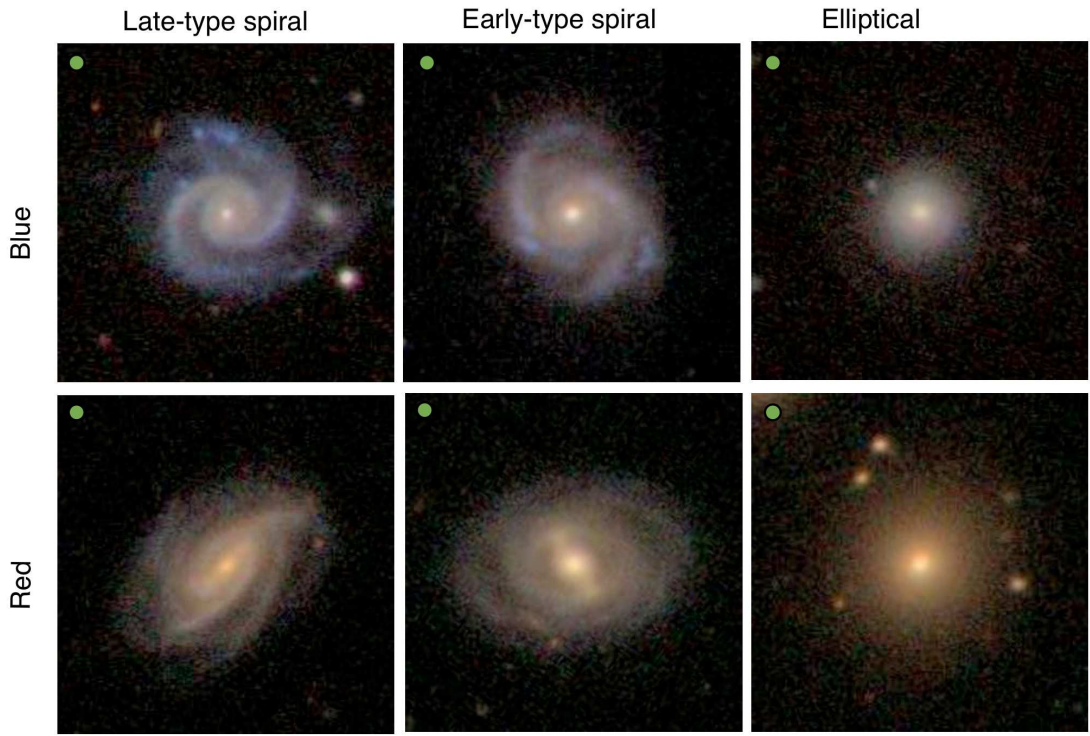}
\vspace{-0.5cm}
\caption{One example from each of the populations studied in this paper. To best showcase their morphology, these examples have some of the largest angular sizes in the samples; panels are $1\times1$ arcmin sq. In each panel the green circle shows the size of the 3-arcsecond spectroscopic fibre. The galaxies are: red elliptical - SDSS J153033.29-004833.5, z=0.077; blue elliptical - SDSS J111215.12+604847.8, z=0.036; red early-type spiral - SDSS J093937.98+594728.5; z=0.048, blue early-type spiral - SDSS J112816.39+005328.8, z=0.040; red late-type spiral - SDSS J144349.31+015533.5, z=0.079; blue late-type spiral - SDSS  J133406.07+014516.0, z=0.032.} 
\label{fig:stamps}
\end{center}
\end{figure*}

These selections result in samples of 13959 red ellipticals, 381 blue ellipticals, 5139 blue late-type spirals, 294 red late-type spirals, and finally 1144 blue early-type spirals, and 1265 red early-type spirals. Recall that all spirals are face-on and have visible arms and therefore our six galaxy samples do not represent the full galaxy population. E.g., we make no statement on edge-on galaxies of either colour, or featureless disk-like galaxies. These numbers also exclude any galaxies which uncertain classifications from Galaxy Zoo (ie. $p_{el}<0.8$ and $p_{sp}<0.8$). The red and blue late-type spiral samples ate identical to those used in \cite{MastersEtAl10}. Fig.~\ref{fig:stamps} shows an example of each galaxy type.

\begin{figure*}
\begin{center}
\includegraphics[angle=90, width=6.2in]{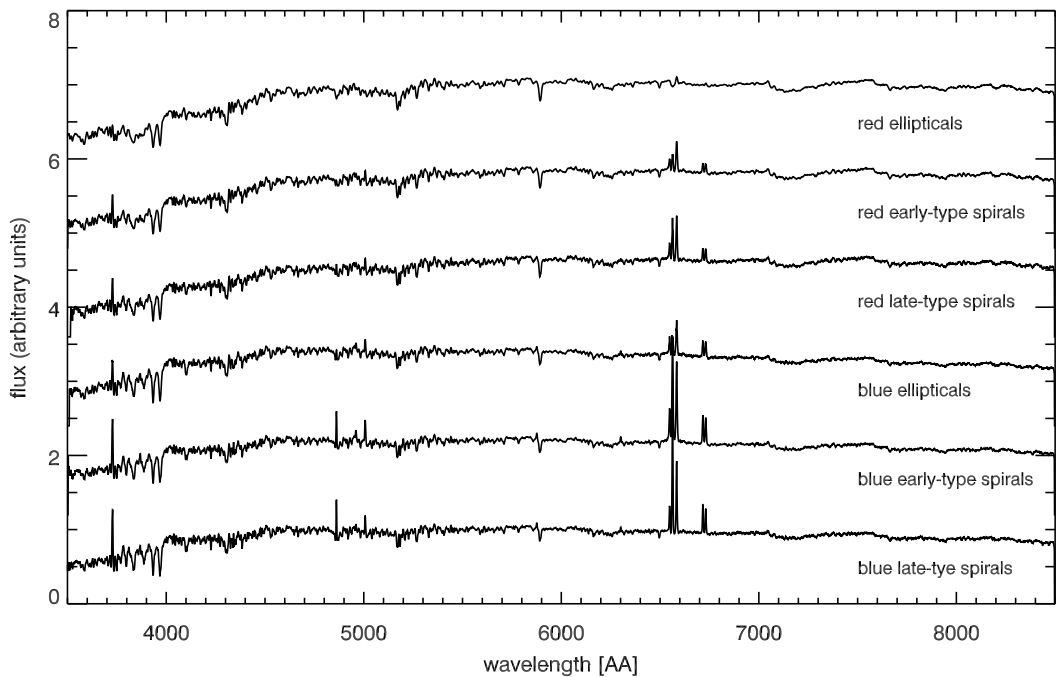}
\caption{The mean spectra of each of the six samples. All fluxes are offset to help visualise differences in emission and absorption features, and from top to bottom the spectra are ordered in median $g-r$ colour (with reddest at the top).} 
\label{fig:stacked_spectra}
\end{center}
\end{figure*}

Although we will refer to red ellipticals and red late/early-type spirals, we note that they do not share the same colour properties: red elliptical are redder than red late- or early-type spirals. The same is true for our blue samples: blue late-type spirals are bluer than blue ellipticals or blue early-type spirals. In $g-r$ it remains true that red late-type spirals are redder than blue ellipticals (the overlap of the red and blue distributions in this plot is simply due to the magnitude dependence of the colour cut that defines the samples), but this separation becomes blurred in $u-r$. In this case, morphology alone separates these two ``green-valley'' populations.

Fig.~\ref{fig:stacked_spectra} shows the stacked spectra of the four samples. Red ellipticals show only very weak emission lines, but all other samples have visible [OII], H$\beta$, [OIII], H$\alpha$, [NII] and [SII] emission. As expected, blue late-type spirals show the strongest lines, dominated by recent by star-formation.  Note that emission lines are {\em not} considered by VESPA, which fits only absorption features and the shape of the continuum.
The purpose of this paper is to turn these observed differences in the spectra of the galaxies into as detailed physical parameters as the data and modelling allow. Whereas these differences appear small to the eye, we will see they are sufficient to reveal very different evolutionary scenarios behind each population of galaxies.

\section{Method}\label{sec:method}

We cross match the samples described above with the public database of star-formation and metallicity histories of \cite{TojeiroEtAl09}; we find matches for over 90\% of the galaxies. The missing 10\% had too poor spectral quality to be analysed reliably (either because too many pixels were masked due to observational issues, or the the surface brightness of the galaxy was too low; these 10\% were not biased in any visible way in stellar mass or redshift). The database provides the stellar mass formed as a function of lookback time, the metallicity as a function of lookback time, and the dust content of each galaxy as recovered with the full spectral-fitting code VESPA \citep{TojeiroEtAl07}. In short, VESPA splits the lookback time of a galaxy in bins that span different ages, and finds the best-fitting combination (in a least-squares sense) of single stellar populations of different metallicities to the observed spectra. The number of age bins is free to change according to the quality of the data, as is their width, and the metallicity is freely allowed to vary as a function of age. Additionally, VESPA fits for dust by using the mixed-slab model of \cite{CharlotFall00}, and up to two optical-depth values: one for stars exclusively younger than 30 Myr (considered still to be enclosed within their birth dust cloud) and one applied to stars of all ages. In this paper we use a one-parameter dust model only, intrinsically assuming the same dust extinction applies to stars of all ages.

\begin{figure}
\begin{center}
\includegraphics[width=6in]{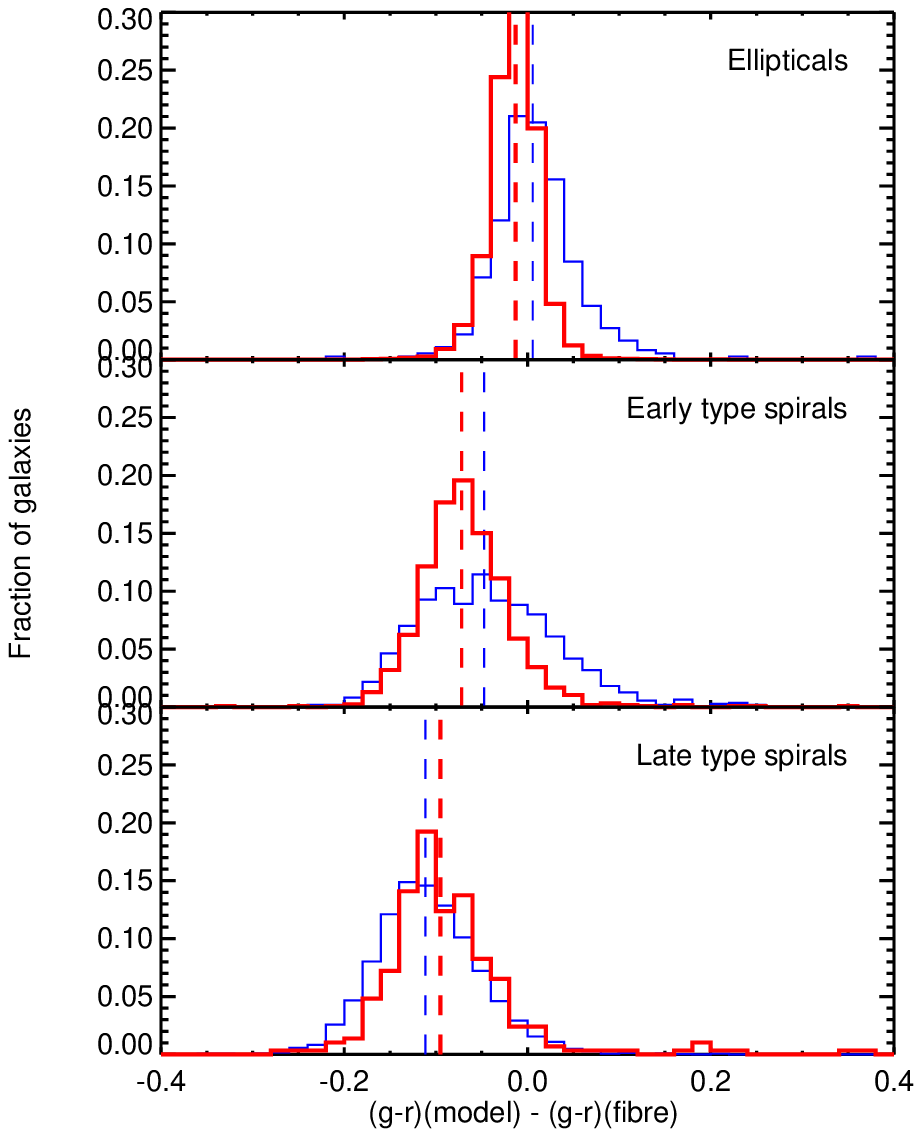}
\caption{The difference in $g-r$ colour using fibre and model photometry for red and blue galaxies (thick and thinner lines, respectively) in all three morphological samples. A negative value corresponds to a redder colour within the 3 arcsec fibre aperture. This figure shows clearly the typical colour gradient of spiral galaxies, which is shown to be steepest for blue late-type spirals. Ellipticals galaxies show no significant offset from a zero mean, although the distribution of red ellipticals is slightly skewed towards negative values.} 
\label{fig:gmr_fibercorr}
\end{center}
\end{figure}

The parameters recovered by VESPA unavoidably refer only to the region of the galaxy that falls within the 3" fibre. For a low-redshift sample such as the one we use here, this effect is important - our results refer only to the central parts of the galaxies in our samples, and must be interpreted within that context. This is particularly relevant for late-type galaxies, where gradients in star-formation and chemical content are more pronounced. Fig.~\ref{fig:gmr_fibercorr} shows the difference in $g-r$ colour from model magnitudes (integrated photometry) and fibre magnitudes (from flux within a 3 arcsec aperture). The colour gradient is shallow for elliptical galaxies, and we observe no significant offset in the median colour difference (0.005 and -0.013 magnitudes for blue and red ellipticals respectively). The distribution in colour difference for red ellipticals is, however, slightly tilted towards negative values (indicating a redder colour within the fibre aperture). As expected, spiral galaxies show a significant colour gradient, and they display a redder colour within the fibre aperture. The gradient is steepest for blue late-type spirals.

Total stellar masses are scaled up from the fibre aperture using the petrosian $z-$band magnitude (see \citealt{TojeiroEtAl09} for more details), and ignore the colour gradients displayed in Fig.~\ref{fig:gmr_fibercorr}. As the total stellar budget is dominated by old stellar populations in all samples, which are the dominant contributors to $z-$band flux, gradients in blue colours such as $g-r$ are expected to have a minor effect in our correction for the total stellar mass. The median colour differences between model and fibre magnitudes in $i-z$ are much smaller, with median values typically half of what is seen in $g-r$.

\subsection{Stellar population models}

Stellar population models are the building blocks of full spectral fitting analyses. They provide the spectral signature of a single stellar population (SSP) of a given age and metallicity. To produce an SSP one must assume an initial mass function that describes the stellar mass distribution of an SSP, stellar tracks that model the different stages of stellar evolution and spectral libraries that model or empirically sample the spectral output of different types of stars. Each of these three main components is important, and affects the predicted spectrum of a single population of stars of a fixed age and metallicity.

Different authors have produced flexible and publicly available codes or sets of SSP models. They differ in one or more of the three components above, and these differences can have significant impact on science analyses (see e.g. \citealt{KolevaEtAl08,TojeiroEtAl09, ConroyAndGunn10, MacArthurEtAl10,TojeiroEtAl11}). While it is important to be aware of the limitations of the SSP modelling, a number of robust conclusions can still be derived, which are not affected by these uncertainties.

To help answer some questions regarding SSP systematics, the VESPA database was published using two sets of SSP models \citep{TojeiroEtAl09} and is in the process of being updated with more data and models (Tojeiro et al. in prep). For the current work we use only publicly available results, using the SSP models of \cite{Maraston05} (M05) and \cite{BruzualEtCharlot03} (BC03).

With the BC03 models we adopt a Chabrier initial mass
function \citep{Chabrier03} and the Padova 1994 evolutionary tracks
\citep{AlongiEtAl93, BressanEtAl93, FagottoEtAl94a,
FagottoeEtAl94b, GirardiEtAl96}. We interpolate metallicities using tabulated values at: $Z=$ 0.0004, 0.004, 0.008, 0.02 and 0.05. We use the red horizontal branch models of \cite{Maraston05} with a Kroupa initial mass function \citep{Kroupa07}. Models are supplied at metallicities of $Z=$ 0.0004, 0.01, 0.02 and 0.04. Both models are normalized to $1M_\odot$ at $t=0$. M05 models are based on the  BaSeL spectral library \citep{LejeuneEtAl97,LejeuneEtAl98} and on empirical C,O stars \citep{LanconWood00} for the TP-AGB phase; BC03 models are based on the STELIB spectral library \citep{LeBorgneEtAl03} in the range from 3200\AA\ to 9500\AA\ and in the BaSeL 3.1 library outwith this range. Note that the Chabrier and Kroupa IMFs do not differ significantly from one another.

Previous comparisons of these two sets of models have consistently shown that M05 give more physically motivated solutions in a variety of settings (see \citealt{TojeiroEtAl09, TojeiroEtAl11}, where BC03 models result in unphysical dips and bumps in the average star-formation histories of galaxies, and unphysical metallicity evolution trends in the populations of luminous red galaxies). In this paper we will therefore focus our interpretation on the results obtained using the M05 models, but replicate all plots using BC03 in Appendix \ref{sec:appendix_BC03}. This enables the effects of differences between SSP models to be observed, and should help with comparing our results to previous work (which is often based on the BC03 models). 

\begin{figure}
\begin{center}
\includegraphics[width=2.5in,angle=90]{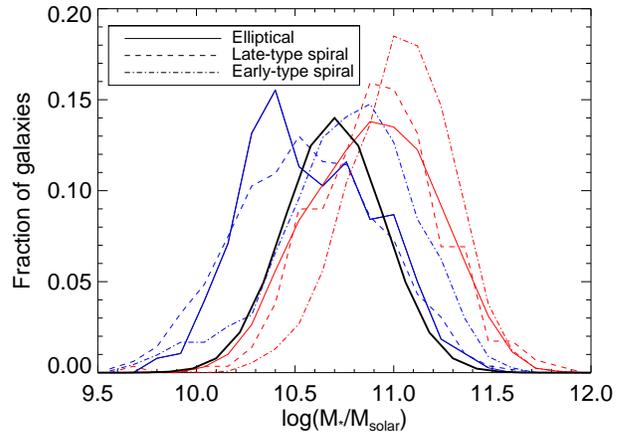}
\caption{The normalised distribution of stellar masses for the six samples of galaxies in this paper (red galaxies in red, or the three rightmost curves; blue galaxies in blue, or the three leftmost curves). The black thick curve shows the stellar mass distribution to which all samples are weighted, peaking and centred at  $M_* \sim 10^{10.7} M_\odot$; this allows us to eliminate stellar mass as a driver of physical differences we observe between the four samples of galaxies.} 
\label{fig:stellar_masses}
\end{center}
\end{figure} 

\begin{figure*}
\begin{center}
\includegraphics[width=7in]{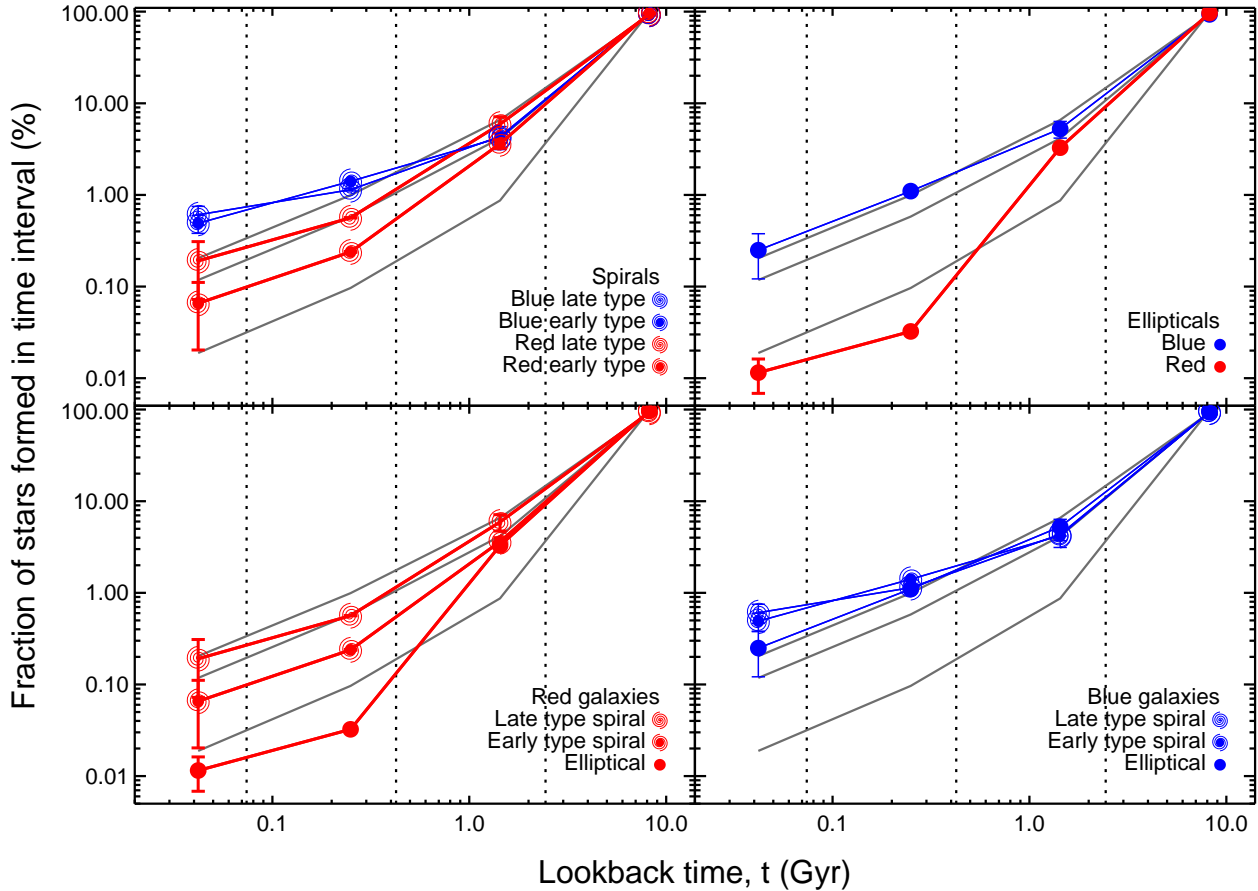}
\caption{The average star formation fraction as a function of lookback time (i.e., a star-formation history) for the six samples studies in this paper (red and blue galaxies in thicker and thinner lines, respectively), computed in four age bins (defined in each panel by the dotted vertical lines) The error bars show the error on the mean. All values are given in Table \ref{tab:SFFs}. For reference, in all panels the grey lines show a star formation history of the form SFR $\propto \exp(\tau_e t$), where $t$ is the lookback time and, from top to bottom, $\tau_e$ is 0.1, 0.15 and 0.3 Gyr$^{-1}$  - see Section~\ref{sec:interpretation} for more details.} 
\label{fig:SFF_M05}
\end{center}
\end{figure*}

\section{Results}\label{sec:results}

\subsection{Stellar masses and star formation histories}

VESPA recovers the amount of star formation in up to 16 age bins, logarithmically spaced between 0.002 Gyr and the age of the Universe (taken here to be 13.7 Gyr, \citealt{KomatsuEtAl11}). All ages refer to the rest-frame of the galaxy. We compute the present-day stellar mass in each galaxy by adding up the stellar mass formed in each age bin, taking into account stellar mass loss to the inter-stellar medium due to stellar evolution (given explicitly by the stellar population models). 

We show the distribution of stellar masses for the six samples in Fig.~\ref{fig:stellar_masses}: as shown in \cite{MastersEtAl10}, red galaxies tend to be more massive, independently of morphology. Blue early-type spirals, however, are more massive than any of the other blue samples. Early-type spirals (red or blue) are very rare at low stellar masses, which explains this trend - note that red early-type spirals are on average slightly more massive than blue early-type spirals. In order to retain as much signal as possible and yet compare samples that are equivalent in terms of stellar mass, we weight each galaxy sample to the weighted stellar-mass distribution shown by the black line in Fig.~\ref{fig:stellar_masses}. The weights are computed by galaxy, and given by the ratio of the target distribution of $\log M_*/M_\odot$ (shown in the black line of Fig.~\ref{fig:stellar_masses}) to a Gaussian fit to each of the population's distributions, at the stellar mass of each galaxy. Weights are normalised to add to unity for each galaxy population. In other words, our results are normalised for galaxy populations with consistent stellar mass distributions, with an effective stellar mass of $M_* \sim 10^{10.7}M_\odot$.

\begin{table*}
 \caption{The mean star-formation fraction (SFF) in each age bin for the six galaxy samples, as displayed in Fig.~\ref{fig:SFF_M05}. The $1\sigma$ column gives the standard error on the mean for each bin. The star-formation fractions and $1\sigma$ errors are given in units of $10^{-3}$. To ease comparison $\alpha$ is the ratio of each star-formation fraction to that of the red ellipticals.    }
\begin{tabular}{|l|c|c|c|c|c|c|c|c|c|c|c|c|c|c|c|}
\hline
  		& \multicolumn{3}{|c|}{ $0.01 - 0.074$ Gyr} &  \multicolumn{4}{|c|}{$0.074 - 0.425$ Gyr} &  \multicolumn{4}{|c|}{ $0.425 - 2.44 $ Gyr} &  \multicolumn{4}{|c|}{$2.44 - 13.7$ Gyr} \\ \hline
 	                        & SFF     	& $1\sigma$  	& $\alpha$	&& SFF     	& $1\sigma$	& $\alpha$	&&   SFF	& $1\sigma$	& $\alpha$		&& SFF    	& $1\sigma$	& $\alpha$\\
Red ellipticals       & $0.11$ 	& $0.047$		& 1			&& $0.32$& $0.0052$	&  1			&&  33		& $1.0$		& 1			&& 966	& 2.89 		& 1 \\ 
Red ET spirals      & $0.65$	 & $0.45$ 		& 5.9    		&& $2.4 $ 	& $0.023$		& 7.5      		&& 36		& 3.8   		& 1.1         		&& 960	& 8.40   		& 0.96 \\
Red LT spirals      & $1.9  $	 & $1.18$		& 17.3    		&& $5.6 $ 	& $0.0097$	& 17.5      		&& 59		& 12   		& 1.8  		&& 933	& 18.7    		& 0.96 \\
Blue ellipticals      & $ 2.5$	& $1.3 $		& 22.7    		&& $11  $ & $0.30$		& 34.4      		&&  52		& 11  		& 1.6        		&& 934	& 17.2    		& 0.96 \\
Blue ET spirals     & $ 4.9$	&  $1.1$		& 44.5  		&& $14  $ & $0.14$		& 43.7     		&& 42		& $5.2$  		& 1.3        		&& 938	& 9.20    		& 0.97\\ 
Blue LT spirals     & $ 6.1 $	&  $1.4$		& 55.5  		&& $11  $ & $0.34$		& 34.4     		&& 43		& $12$  		& 1.3        		&& 939	& 19.3  		& 0.97\\ \hline
 \end{tabular}
 \label{tab:SFFs}
 \end{table*}
 
To compute the mean star formation history of each galaxy sample we 
re-bin the original 16 age bins into four wide age bins. Using just four age bins minimises the correlations between adjacent bins of star-formation, whilst still maintaining significant information about the star formation history of each sample of galaxies. Note that we will work with mean star-formation {\em fractions} - i.e., all star-formation histories shown add up to unity over all cosmic time.

Fig. ~\ref{fig:SFF_M05} shows the star formation histories of the six samples, and we give the average fraction of stellar mass formed in each age bin in Table~\ref{tab:SFFs}. We can make the following observations:
\begin{enumerate}
\item Red classical ellipticals consistently show the most steeply decaying star-formation history, and extremely low star formation fraction ($<1\%$) in the two youngest bins ($<0.5$Gyr).
\item Red late-type spirals show a lower amount of recent star-formation than any of the blue samples, but only by up to a factor of 3, whereas they show around {\it 17 times more recent star-formation than red ellipticals}. 
\item Red and blue late-type spirals' histories depart from each other at young ages ($<0.5$ Gyr).
\item The star-formation histories of blue late- and early-type spirals are very similar.
\item Blue spirals and blue elliptical galaxies have remarkably similar star-formation histories, the exception being the youngest bin ($<100$ Myr), where blue spirals show a larger a star-formation fraction by a factor of two.
\item Red early-type spirals have around 3 times less recent star-formation than red late-type spirals.
\end{enumerate}

\begin{figure}
\begin{center}
\includegraphics[width=4in, angle=90]{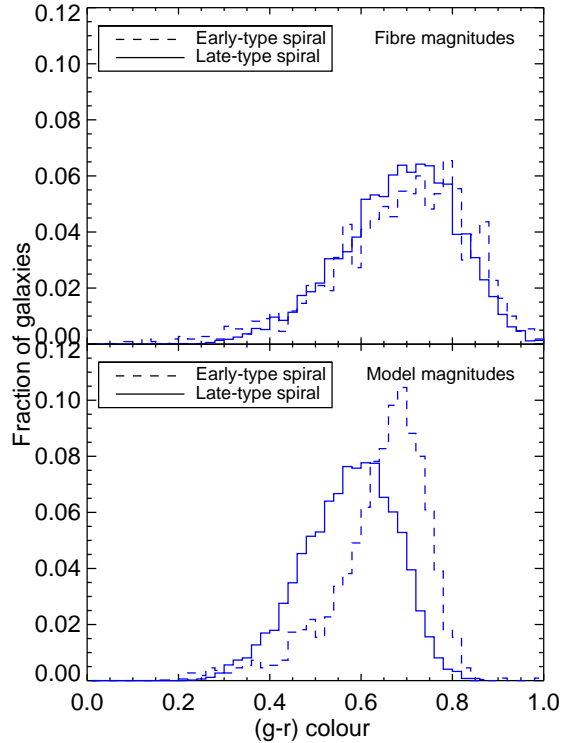}
\caption{The observed $g-r$ colours within the 3" fibre-aperture (top panel) and computed using integrated photometry (bottom) for blue early- and late-type spiral galaxies. The fibre colour is nearly identical for the two samples of spiral galaxies, indicating the difference seen in integrated photometry colour in Fig.~\ref{fig:grplot_bulgy} is dominated by the outer regions of the galaxies. Coupled with the results in Fig.~\ref{fig:SFF_M05}, these results suggest that the main difference between early- and late-type spirals lies in the {\em extent} of the bulge component, rather than in its stellar composition. } 
\label{fig:spiral_colours}
\end{center}
\end{figure}

Point (iv) seems at odds with Fig.~\ref{fig:grplot_bulgy}, which shows that blue early- and late-type spirals have different distributions of $g-r$ colours. However, we note that the fibre-aperture colours of the two blue spiral samples are identical - i.e., the difference in the colours from integrated photometry seen in Fig.~\ref{fig:grplot_bulgy} is dominated by the outer regions of the galaxies, as seen in Fig.~\ref{fig:spiral_colours}. This gives a scenario where the main difference between the two samples is the {\em extent} of the bulge component, not its stellar content. The same cannot be said of the red late- and early-type spirals. Both their fibre colour and spectra suggest that early-type red spiral galaxies have weaker recent star-formation (by roughly a factor of 3) than late-type red spirals.

We next look at the metallicity content of each of the four samples.

\subsection{Stellar Metallicity}\label{sec:metallicity}
\begin{figure}
\begin{center}
\includegraphics[width=6in]{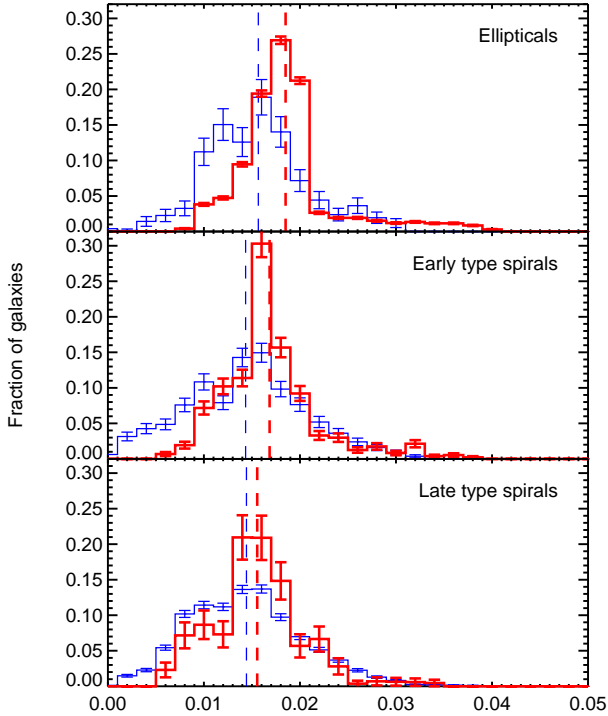}
\caption{The distribution of the mass-weighted stellar metallicity for red and blue galaxies (thicker and thinner lines, respectively) for all three morphological samples, using M05 modelling. The vertical dashed line shows the mean value of each distribution. The error bars are Poisson errors. } 
\label{fig:metallicity_M05}
\end{center}
\end{figure}

VESPA returns a stellar metallicity for each age bin. Throughout his paper we quote metallicity in units of the mass fraction of metals with respect to Hydrogen; in these units solar metallicity is $Z_\odot = 0.02$.  Whereas in principle VESPA gives a metallicity history for each galaxy type, the data quality is not good enough to yield reliable results - \cite{TojeiroEtAl09} showed in particular how metallicity was virtually unconstrained in age bins populated with small amounts of stellar mass, but was reliably recovered if one only considered the two most significant bins in terms of mass. We therefore compute a mass-weighted metallicity per galaxy, which automatically deals with this issue and retains most of the useful information. Using a mass-weighted metallicity also results in a continuous distribution of metallicity values - for age bins populated only with small amounts of stellar mass, VESPA is more likely to return a metallicity at the exact values provided by the SSP models, resulting in an artificially discrete distribution of metallicities. 

Fig.~\ref{fig:metallicity_M05} shows the distribution of mass-weighted metallicities for the six galaxy samples. We can make the following statements:
\begin{figure*}
\begin{center}
\includegraphics[angle=90,width=5in]{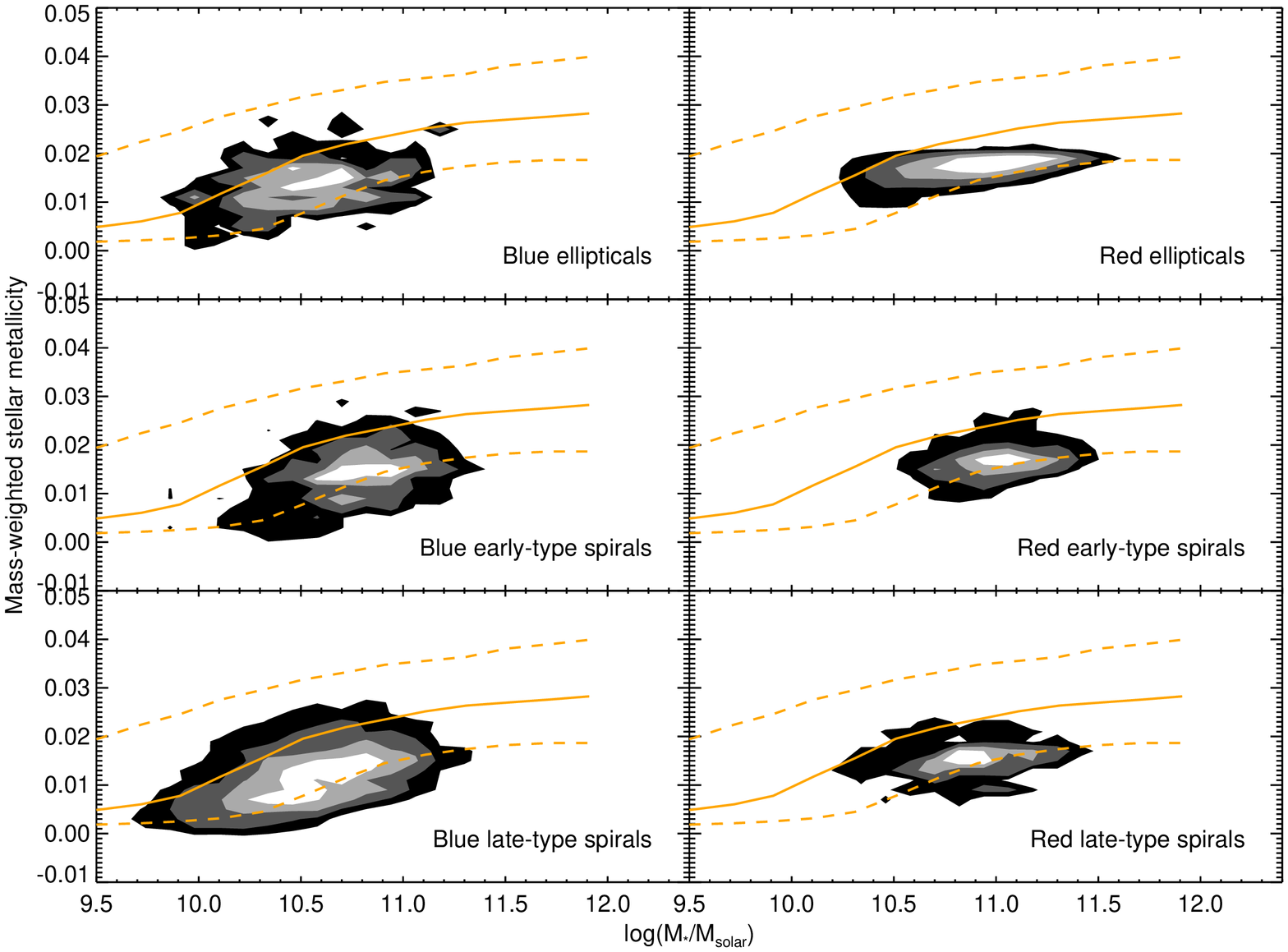}
\caption{Stellar mass vs metallicity relation for the six samples. Left-hand side shows blue galaxies, and right-hand side shows red galaxies. The orange solid and dashed lines show the stellar mass - metallicity relation measured by \protect\cite{GallazziEtAl05}. The offset in metallicity between our mass vs metallicity relationship and that of Gallazzi et al. is explained by the stellar population models - Gallazzi et al. use BC03, and this offset is much reduced when we use the same set of models (see Fig.~\ref{fig:2D_metallicity_BC03}).} 
\label{fig:2D_metallicity_M05}
\end{center}
\end{figure*}

\begin{figure}
\begin{center}
\includegraphics[width=6in]{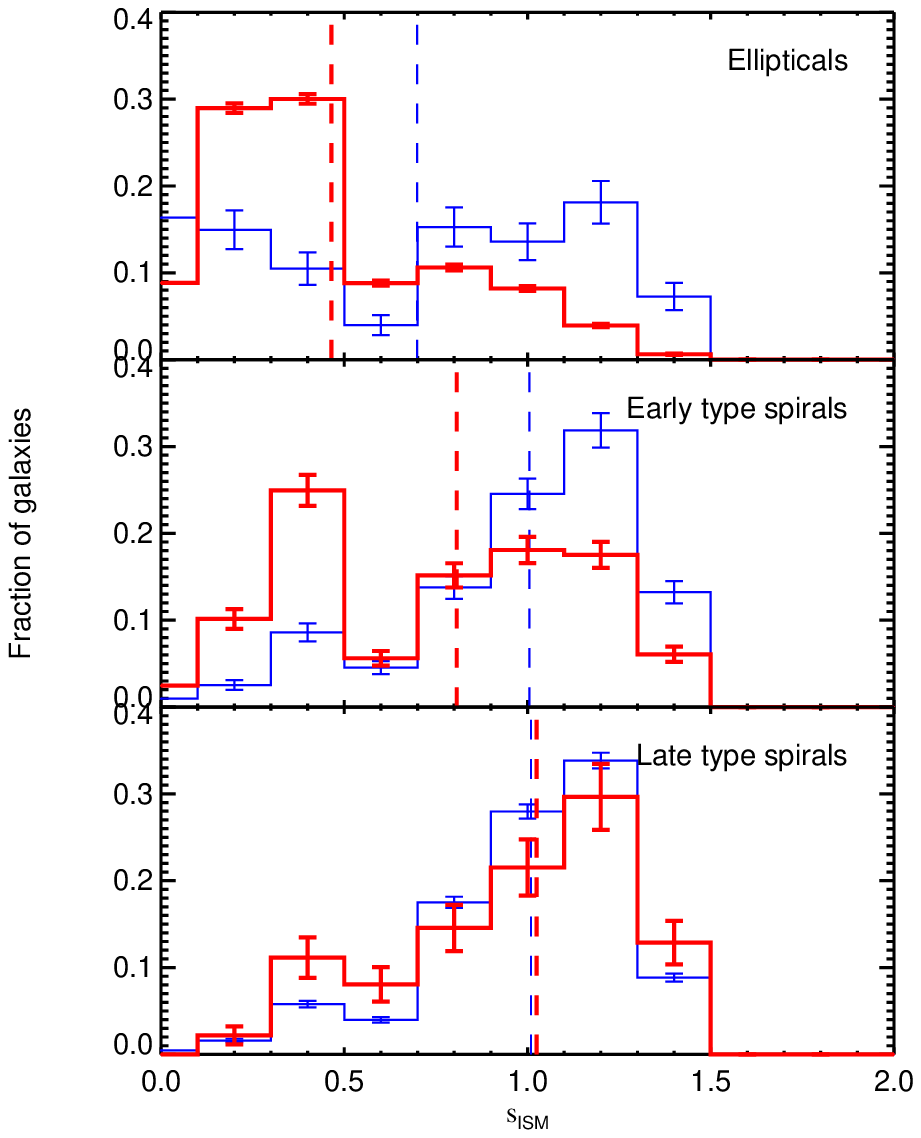}
\caption{The distribution of the optical depth $\tau_{ISM}$ for red and blue galaxies (thicker and thinner lines, respectively) for all three morphological samples, using M05 modelling.  The vertical dashed line shows the mean value of each distribution. The error bars are Poisson errors. } 
\label{fig:dust_M05}
\end{center}
\end{figure}

\begin{enumerate}
\item Blue galaxies are on average marginally more metal poor than red galaxies. This statement is independent of morphology, but more significant for elliptical galaxies.
\item Red ellipticals are marginally more metal rich than red spirals. 
\item Red galaxies show a smaller spread in metallicity than blue galaxies, except in the case of late-type spirals.
\item There is no significant difference between metallicity distribution of early- and late-type spirals
\end{enumerate}

The relationship between stellar mass and metallicity provides clues and constraints on the chemical enrichment processes within the galaxies. Now dropping the mass-weighting scheme used elsewhere in this paper, we show the mass-metallicity relation for the six samples in Fig.~\ref{fig:2D_metallicity_M05} . The mass-metallicity relation seems steeper for blue galaxies independently of morphology, but this is a reflection of the different range in stellar mass - the mass-metallicity relation is known to be steeper at lower masses (see \citealt{GallazziEtAl05}, and the corresponding yellow line in both panels). There is an offset in metallicity between our results and the mass-metallicity relation found in \cite{GallazziEtAl05}. This is explained by the different stellar population models in the two analyses - Gallazzi et al. use BC03. See Fig.~\ref{fig:2D_metallicity_BC03} for a comparison with the two sets of results using the BC03 models - the agreement is visibly good. 
Given the intrinsic scatter and the limited sizes of the samples, we observe no significant difference in the mass-metallicity relation of the six samples.

\subsection{Dust}

We model dust extinction using the mixed-slab model of \cite{CharlotFall00} and an extinction curve of the form $\tau_{\lambda} = \tau_{ISM} (\lambda / 5500\AA) ^ {-0.7}$, where $\tau_{ISM}$ is the optical depth at a wavelength of $5500 \AA$, and the only free parameter of our dust model. 

Fig.~\ref{fig:dust_M05} shows the distribution of $\tau_{ISM}$ values for the six samples. We can make the following statements: 
\begin{enumerate}
\item On average spiral galaxies are dustier than elliptical galaxies independently of colour. 
\item The dust distribution of red and blue late-type spirals is remarkably similar, indicating that dust is unlikely to be the reason red spirals are red (recall these are face-on spirals only). 
\item  Blue ellipticals are dustier than red ellipticals but they show a bimodal distribution in dust extinction, with a highly obscured population that is present but less common in the population of red ellipticals.
\item Red early-type spirals have a lower average dust extinction than blue late-type spirals ($\tau_{ISM} \approx 0.8$ vs $\tau_{ISM} \approx 1$, respectively). This is due to the presence of a population of objects with very low extinction in the red population, which is not seen in the blue early-type spirals.
\end{enumerate}

In the red population we see a distinct increase of the presence of low-extinction objects with growing bulge size: late-type spirals, selected to have {\tt fracdeV} $ < 0.5$, have $\sim 10\%$ of objects with $\tau_{ISM} < 0.5$, with this fraction increasing just over 30\% in early-type spirals ({\tt fracdeV} $ > 0.5$) and to over 60\% in ellipticals. This result is consistent with larger bulges having a composition with decreasing dust-extinction.

The small number of blue ellipticals (381) makes it difficult to reliably disentangle the dust bimodality observed in the upper panel using other properties. Stellar mass does not seem to be a factor, as the mass-weighting brings little change to the dust distribution. A visual inspection of the objects did not uncover any obvious difference between the blue ellipticals with high and low obscuration, nor does dust have a noticeable dependence on colour gradients.

\subsection{Colours}\label{sec:interpretation}
\begin{figure*}
\begin{center}
\includegraphics[angle=90,width=6.5in]{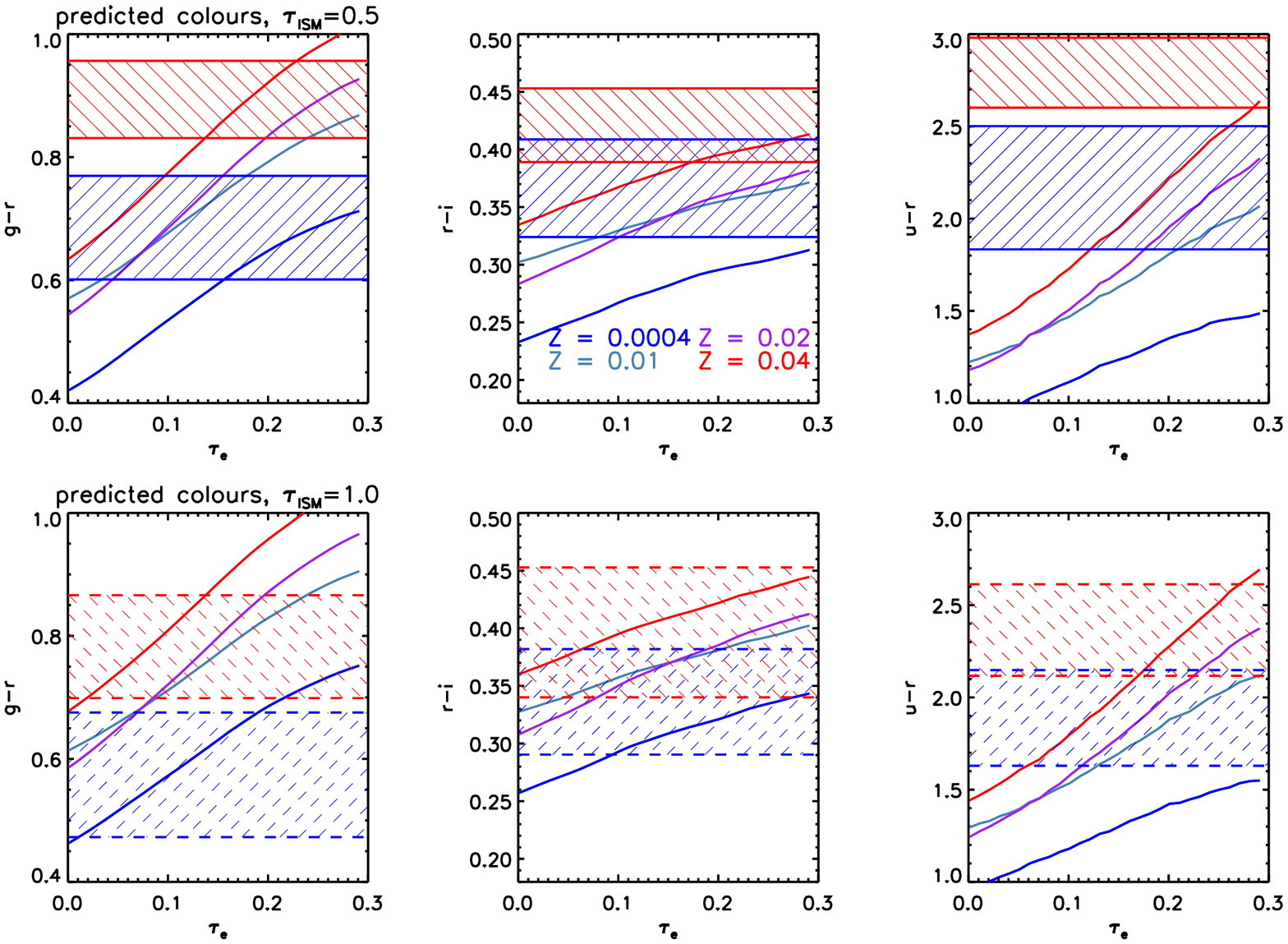}
\caption{Smooth curves show the predicted {\em observed} $g-r$ (left), $r-i$ (centre) and $u-r$ (right) colours of our toy models of star formation for different metallicity values, assuming $\tau_{ISM}=1$ (top row) and $\tau_{ISM}=0.5$ (bottom row), and computed using the  M05 SSP models. The age-metallicity degeneracy is clearly visible in this diagram. The horizontal bands show where the observed colours of the four samples lie (regions within $\pm 1\sigma$ of the mean observed colour). In accordance to the extinction values found for each sample, on the top row we show the observed early-type galaxies, and on the bottom late-type galaxies. Values of $\tau_e$ and $Z$ that overlap with the regions of observed colour are the ones able to describe the samples, assuming our simple toy model. } 
\label{fig:all_colours}
\end{center}
\end{figure*}

The results in the previous Section quantify differences in the physical properties of the six samples of galaxies we consider. To create some meaningful context in this Section we consider how these physical properties affect the observed colours of galaxies. 

Fig.~\ref{fig:all_colours} shows the predicted observed $g-r$, $r-i$ and $u-r$ colours as a function of star-formation history and metallicity, computed using the M05 models. We parametrize the star-formation history with an exponentially decaying form as SFR $\propto \exp(\tau_e t$) where $t$ is the lookback time in Gyr, and which loosely follows the behaviour of our samples. This is too simplistic a model to allow good fits in all cases, but it allows us to study the predicted colours as a function of a single, smooth parameter that roughly bridges the gap between the SFHs of our samples. For reference we plot the star-formation history corresponding to $\tau_e = 0.1, 0.15$ and $0.3$ Gyr$^{-1}$ as the grey lines in Figs.~\ref{fig:SFF_M05}. Blue galaxies have $\tau_e \approx 0.1$, red late-type spirals $\tau_e \approx 0.15$ and red early-types $\tau_e \approx 0.3$.

Focusing first on red and blue ellipticals, the top left panel of Fig.~\ref{fig:all_colours} shows that the expected difference in colour between $\tau_e \sim 0.1$ and $\tau_e \sim 0.3$ is sufficient to explain the observed differences in colour at all metallicities except the very low or very high end. In other words, star-formation alone (and in particular, recent star formation) is sufficient to explain the different classification of these two samples. 

The star-formation histories of blue ($\tau_e \sim 0.1$) and red ($\tau_e \sim 0.15$) late-type spirals are much more similar, but Fig.~\ref{fig:all_colours} shows that even such small differences can yield measurably different observed colours, at least in $g-r$ which the colour that drives our classifications (see bottom left panel). Our toy model suggests that blue early-types are either slightly younger or more metal poor than the fully-fitted SFHs presented in the previous Section: indeed, our toy model poorly fits blue early-types at ages younger than 100 Myr. Nonetheless, the take-home message is that even small differences in recent star-formation rates can visibly impact the colours of the samples, and galaxy colour-classification.

Note that the differences in metallicity measured in Section~\ref{sec:metallicity} are too small to produce visible differences in colour (given the scatter in colour and size of the sample). 

\section{Summary and conclusions}\label{sec:discussion}

We have used VESPA to study the stellar, chemical and dust content of six sample of galaxies, defined according to their colour and morphology. Our findings can be summarised in the following bullet points:

\begin{itemize}
\item Red ellipticals consistently show the most steeply decaying star-formation history, and extremely low star formation fraction ($<1\%$) in the two youngest bins ($<0.5$Gyr). They are the most metal rich of the six samples and display the lowest amount of dust obscuration. These results are consistent with red ellipticals having the reddest rest-frame colours, and with the typical view of early-type galaxies as being old, metal rich and dust-free objects.
\item Red late-type (disk-dominated) spirals show a lower amount of recent star-formation than any of the blue samples, by up to a factor of 3. However, they show around {\it 17 times more recent star-formation than red ellipticals}. Red and blue spirals' star-formation histories depart significantly only at at young ages ($<0.5$ Gyr). The redder colour of red late-type spiral galaxies (when compared to blue late-type spirals) can be explained simply by this small difference in recent star-formation. Red late-type spirals have a similar average metallicity than blue spirals, but a larger scatter in metallicity values.  The dust distribution of blue and red spirals, when observed in face-on orientation, is incredibly similar - both samples are highly obscured compared to ellipticals. 
\item Blue ellipticals have remarkably similar fibre star-formation histories to blue spirals, the exception being the youngest bin ($<100$ Myr), where blue spirals show a larger a star-formation fraction by a factor of two. Blue spirals and ellipticals do not differ significantly in their chemical content, but they display largely different dust distributions - in the case of blue ellipticals this seems bimodal, with one population of dust-free objects (similar to red ellipticals) and one population of highly obscured objects (similar to blue spirals). We could not disentangle this bimodality using stellar mass, colour gradients, metallicity or visual inspection.
\item Blue late-type (disk-dominated) spirals have the flattest star-formation history, and the most recent star-formation. They are on average the most metal poor of the four populations, and are on average more dust-obscured than blue ellipticals. 
\item Red early-type (bulge-dominated) spirals have 3 times less recent star-formation than red late-type spirals, and slightly lower dust extinction: in both cases they sit in between red late-type spiral and red elliptical galaxies. 
\item Blue early-type (bulge-dominated) spirals have identical fibre star-formation histories to blue late-type spirals, in spite of the fact they display different colours in integrated photometry (blue early-type spirals are redder). Their fibre colour is, however, remarkably similar and suggesting that the change in integrated photometry colours is driven by the regions outwith the radius probed by the spectroscopic fibre. I.e., the extent of the bulge in these two samples differs, with it being more extended (but not necessarily of different stellar content) in early-type blue spirals.
\end{itemize}

Our results show that red late-type spiral galaxies are red because of a genuine decline in star-formation history in the last 0.5 Gyr: their star-formation history prior to that point, dust and chemical content is similar to blue late-type spirals. This is consistent with a scenario where red late-type spirals share a common evolutionary path with blue late-type spirals, and whatever mechanism causes a diminishing of their star-formation rate must be gentle enough not to disturb their morphology, nor to substantially impact on the dust and chemical content. According to our modelling, {\it this process must last around 0.5 Gyrs} - if it were significantly shorter (less than 100 Myrs) or longer (more than 1Gyr) we would expect the measured star-formation histories to depart significantly either at a younger or older age bin. Due to our stellar mass weighting scheme, our results effectively refer to a stellar mass $M_* \sim 10^{10.7}M_\odot$, but it remains true that red late-type spirals are on average more massive than blue late-type spirals (see Fig.~\ref{fig:stellar_masses}, and barring any selection effects not considered), suggesting that more massive blue late-type spirals must be more effective at turning red, if they do so at all. This is in agreement with the results of \cite{SkibbaEtAl09c, BamfordEtAl09}, who -- albeit working on a sample of red spirals that included both highly inclined and bulge-dominated objects -- found that red spirals are predominantly found in denser environments, which in turn is correlated with stellar (and halo) mass. Additionally they found that, at fixed stellar mass, environment drives colour more strongly than it drives morphology. They argue that colour changes due to environmental effects (``strangulation", ``harassment", etc) affect colour in a shorter time-scale than they affect morphology; this is consistent with the results in this paper.

Whereas naively one could constrain the transition period between blue late-type spiral and blue elliptical galaxies using an argument similar to the above, we note that this comparison is hindered by fibre-aperture effects, which are much more pronounced for the spiral population relatively to the ellipticals (Fig.~\ref{fig:gmr_fibercorr}). The global recent-intermediate star-formation of late-type spiral is likely larger than what we measure through the fibre, making it harder to discern how much of a history the two populations truly share. Whereas the two share a similar chemical content, they have sufficiently different dust contents to suggest that blue ellipticals are likely not the result of a direct transition from blue late-type spirals, or at least not exclusively. Indeed, it may be that the path that leads to the formation of blue ellipticals is not unique - the broad (marginally bimodal) dust distribution of this population leaves room for a scenario where blue ellipticals can come to be as rejuvenated red ellipticals {\it or} as truncated blue late- or early-type spirals (with this process leading to a change of morphology via something like a merger, but not truncating the star-formation completely - see also \citealt{SchawinskiEtAl09}).

It is interesting that red elliptical galaxies lie isolated within the parameter space we explore - they uniformly show low dust extinction values, have extremely low recent-intermediate star-formation rates, and a very uniform distribution in metal content, that peaks at the largest value of all the populations we study. This is of course consistent with the well-known lower scatter of the red-sequence when compared to the blue cloud; our work simply puts this result in terms of physical properties, rather than observed colours and magnitudes. From an evolutionary point of view, it suggests they are the end result of an evolutionary path that stabilised over at least 3 Gyrs (and likely longer, but which we cannot probe with the time resolution used here), and that any mild events of star-formation since then have not disrupted the galaxy's morphology, dust or chemical composition in a significant way. 

Finally, red early-type spiral galaxies lie between red late-type spirals and red ellipticals in terms of recent-to-intermediate star-formation rate, dust content and morphology, making a scenario where upon they mark a transitional period that precedes red ellipticals very plausible. We note they make better candidates for such an evolutionary link than red late-type spirals, which show a star-formation history and dust content much more similar to blue late-type spirals and very unlike red ellipticals. This is in agreement with \cite{MastersEtAl10} who, based on the fraction of galaxies with bars, suggest that early-type red spirals are more likely to evolve into red ellipticals than late-type red spirals.

A realistic galaxy evolution description likely involves a more complex link between these six populations of galaxies - a detailed study using the evolution of their number density, coupled with detailed fossil-record star-formation histories (that allow linking populations across different redshifts - see e.g. \citealt{TojeiroEtAl11b}) would provide further insight onto how all these links come together to produce the observed Universe; this is the topic of future work.  The colour classification in this paper is based on model-magnitudes, but we note that qualitatively all of the conclusions we present in this section are robust to a change from model to petrosian magnitudes.


\section{Acknowledgments}

We would like to thank the anonymous referee for useful comments that led to visible improvements to this paper.WJP and RT acknowledge financial support from the European Research Council under the European CommunityÕs Seventh Framework Programme (FP7 2007-2013) ERC grant agreement n. 202686. WJP is also grateful for support from the UK Science and Technology Facilities Council through the grant ST/I001204/1. JR is thankful for a Ogden Trust summer internship. 
KLM acknowledges funding from The Leverhulme Trust as a 2010 Early Career Fellow. SPB gratefully acknowledges an STFC Advanced Fellowship.

This publication has been made possible by the participation of more than 160 000 volunteers in the first phase of the Galaxy Zoo project. Their contributions are individually acknowledged at http://www.galaxyzoo.org/volunteers.

Funding for the SDSS and SDSS-II has been provided by the Alfred
    P. Sloan Foundation, the Participating Institutions, the National
    Science Foundation, the U.S. Department of Energy, the National
    Aeronautics and Space Administration, the Japanese Monbukagakusho,
    the Max Planck Society, and the Higher Education Funding Council
    for England. The SDSS Web Site is http://www.sdss.org/. The SDSS is managed by the Astrophysical Research Consortium for the Participating Institutions. The Participating Institutions are the American Museum of Natural History, Astrophysical Institute Potsdam, University of Basel, University of Cambridge, Case Western Reserve University, University of Chicago, Drexel University, Fermilab, the Institute for Advanced Study, the Japan Participation Group, Johns Hopkins University, the Joint Institute for Nuclear Astrophysics, the Kavli Institute for Particle Astrophysics and Cosmology, the Korean Scientist Group, the Chinese Academy of Sciences (LAMOST), Los Alamos National Laboratory, the Max-Planck-Institute for Astronomy (MPIA), the Max-Planck-Institute for Astrophysics (MPA), New Mexico State University, Ohio State University, University of Pittsburgh, University of Portsmouth, Princeton University, the United States Naval Observatory, and the University of Washington.

\vspace{-3mm}
\bibliographystyle{mn2e}
\bibliography{/users/ritatojeiro/Dropbox/PAPER/my_bibliography}

\begin{thebibliography}{}

\bibitem[\protect\citeauthoryear{{Abazajian} et~al.,}{{Abazajian}
  et~al.}{2009}]{AbazajianEtAl09}
{Abazajian} K.~N.,  et~al., 2009, \apjs, 182, 543

\bibitem[\protect\citeauthoryear{{Alongi}, {Bertelli}, {Bressan}, {Chiosi},
  {Fagotto}, {Greggio} \& {Nasi}}{{Alongi} et~al.}{1993}]{AlongiEtAl93}
{Alongi} M.,  {Bertelli} G.,  {Bressan} A.,  {Chiosi} C.,  {Fagotto} F.,
  {Greggio} L.,    {Nasi} E.,  1993, \aaps, 97, 851

\bibitem[\protect\citeauthoryear{{Baldry}, {Balogh}, {Bower}, {Glazebrook},
  {Nichol}, {Bamford} \& {Budavari}}{{Baldry} et~al.}{2006}]{BaldryEtAl06}
{Baldry} I.~K.,  {Balogh} M.~L.,  {Bower} R.~G.,  {Glazebrook} K.,  {Nichol}
  R.~C.,  {Bamford} S.~P.,    {Budavari} T.,  2006, \mnras, 373, 469

\bibitem[\protect\citeauthoryear{{Bamford} et~al.,}{{Bamford}
  et~al.}{2009}]{BamfordEtAl09}
{Bamford} S.~P.,  et~al., 2009, \mnras, 393, 1324

\bibitem[\protect\citeauthoryear{{Bell}, {Zheng}, {Papovich}, {Borch}, {Wolf}
  \& {Meisenheimer}}{{Bell} et~al.}{2007}]{BellEtAl07}
{Bell} E.~F.,  {Zheng} X.~Z.,  {Papovich} C.,  {Borch} A.,  {Wolf} C.,
  {Meisenheimer} K.,  2007, \apj, 663, 834

\bibitem[\protect\citeauthoryear{{Bernardi}, {Shankar}, {Hyde}, {Mei},
  {Marulli} \& {Sheth}}{{Bernardi} et~al.}{2010}]{BernardiEtAl10}
{Bernardi} M.,  {Shankar} F.,  {Hyde} J.~B.,  {Mei} S.,  {Marulli} F.,
  {Sheth} R.~K.,  2010, \mnras, 404, 2087

\bibitem[\protect\citeauthoryear{{Blanton} et~al.,}{{Blanton}
  et~al.}{2003}]{BlantonEtAl03}
{Blanton} M.~R.,  et~al., 2003, \apj, 594, 186

\bibitem[\protect\citeauthoryear{{Bressan}, {Fagotto}, {Bertelli} \&
  {Chiosi}}{{Bressan} et~al.}{1993}]{BressanEtAl93}
{Bressan} A.,  {Fagotto} F.,  {Bertelli} G.,    {Chiosi} C.,  1993, \aaps, 100,
  647

\bibitem[\protect\citeauthoryear{{Bruzual} \& {Charlot}}{{Bruzual} \&
  {Charlot}}{2003}]{BruzualEtCharlot03}
{Bruzual} G.,  {Charlot} S.,  2003, \mnras, 344, 1000

\bibitem[\protect\citeauthoryear{{Bundy} et~al.,}{{Bundy}
  et~al.}{2006}]{BundyEtAl06}
{Bundy} K.,  et~al., 2006, \apj, 651, 120

\bibitem[\protect\citeauthoryear{{Bundy} et~al.,}{{Bundy}
  et~al.}{2010}]{BundyEtAl10}
{Bundy} K.,  et~al., 2010, \apj, 719, 1969

\bibitem[\protect\citeauthoryear{{Chabrier}}{{Chabrier}}{2003}]{Chabrier03}
{Chabrier} G.,  2003, \pasp, 115, 763

\bibitem[\protect\citeauthoryear{{Charlot} \& {Fall}}{{Charlot} \&
  {Fall}}{2000}]{CharlotFall00}
{Charlot} S.,  {Fall} S.~M.,  2000, \apj, 539, 718

\bibitem[\protect\citeauthoryear{{Cisternas} et~al.,}{{Cisternas}
  et~al.}{2011}]{CisternasEtAl11}
{Cisternas} M.,  et~al., 2011, \apj, 726, 57

\bibitem[\protect\citeauthoryear{{Conroy} \& {Gunn}}{{Conroy} \&
  {Gunn}}{2010}]{ConroyAndGunn10}
{Conroy} C.,  {Gunn} J.~E.,  2010, \apj, 712, 833

\bibitem[\protect\citeauthoryear{{Conselice}}{{Conselice}}{2006}]{Conselice06}
{Conselice} C.~J.,  2006, \apj, 638, 686

\bibitem[\protect\citeauthoryear{{Cortese}}{{Cortese}}{2012}]{Cortese12}
{Cortese} L.,  2012, \aap, 543, A132

\bibitem[\protect\citeauthoryear{{Eisenstein}, {Weinberg}, {Agol}, {Aihara},
  {Allende Prieto}, {Anderson}, {Arns}, {Aubourg}, {Bailey}, {Balbinot} \& et
  al.}{{Eisenstein} et~al.}{2011}]{EisensteinEtAl11}
{Eisenstein} D.~J.,  {Weinberg} D.~H.,  {Agol} E.,  {Aihara} H.,  {Allende
  Prieto} C.,  {Anderson} S.~F.,  {Arns} J.~A.,  {Aubourg} {\'E}.,  {Bailey}
  S.,  {Balbinot} E.,    et al. 2011, \aj, 142, 72

\bibitem[\protect\citeauthoryear{{Fagotto}, {Bressan}, {Bertelli} \&
  {Chiosi}}{{Fagotto} et~al.}{1994a}]{FagottoEtAl94a}
{Fagotto} F.,  {Bressan} A.,  {Bertelli} G.,    {Chiosi} C.,  1994a, \aaps,
  104, 365

\bibitem[\protect\citeauthoryear{{Fagotto}, {Bressan}, {Bertelli} \&
  {Chiosi}}{{Fagotto} et~al.}{1994b}]{FagottoeEtAl94b}
{Fagotto} F.,  {Bressan} A.,  {Bertelli} G.,    {Chiosi} C.,  1994b, \aaps,
  105, 29

\bibitem[\protect\citeauthoryear{{Fukugita} et~al.,}{{Fukugita}
  et~al.}{2007}]{FukugitaEtAl07}
{Fukugita} M.,  et~al., 2007, \aj, 134, 579

\bibitem[\protect\citeauthoryear{{Fukugita}, {Ichikawa}, {Gunn}, {Doi},
  {Shimasaku} \& {Schneider}}{{Fukugita} et~al.}{1996}]{FukugitaEtAl96}
{Fukugita} M.,  {Ichikawa} T.,  {Gunn} J.~E.,  {Doi} M.,  {Shimasaku} K.,
  {Schneider} D.~P.,  1996, \aj, 111, 1748

\bibitem[\protect\citeauthoryear{{Gallazzi}, {Charlot}, {Brinchmann}, {White}
  \& {Tremonti}}{{Gallazzi} et~al.}{2005}]{GallazziEtAl05}
{Gallazzi} A.,  {Charlot} S.,  {Brinchmann} J.,  {White} S.~D.~M.,
  {Tremonti} C.~A.,  2005, \mnras, 362, 41

\bibitem[\protect\citeauthoryear{{Girardi}, {Bressan}, {Chiosi}, {Bertelli} \&
  {Nasi}}{{Girardi} et~al.}{1996}]{GirardiEtAl96}
{Girardi} L.,  {Bressan} A.,  {Chiosi} C.,  {Bertelli} G.,    {Nasi} E.,  1996,
  \aaps, 117, 113

\bibitem[\protect\citeauthoryear{{Gunn} et~al.,}{{Gunn}
  et~al.}{1998}]{GunnEtAl98}
{Gunn} J.~E.,  et~al., 1998, \aj, 116, 3040

\bibitem[\protect\citeauthoryear{{Gunn} et~al.,}{{Gunn}
  et~al.}{2006}]{GunnEtAl06}
{Gunn} J.~E.,  et~al., 2006, \aj, 131, 2332

\bibitem[\protect\citeauthoryear{{Holden}, {van der Wel}, {Rix} \&
  {Franx}}{{Holden} et~al.}{2012}]{HoldenEtAl12}
{Holden} B.~P.,  {van der Wel} A.,  {Rix} H.-W.,    {Franx} M.,  2012, \apj,
  749, 96

\bibitem[\protect\citeauthoryear{{Holmberg}}{{Holmberg}}{1958}]{Holmberg58}
{Holmberg} E.,  1958, Meddelanden fran Lunds Astronomiska Observatorium Serie
  II, 136, 1

\bibitem[\protect\citeauthoryear{{Hubble}}{{Hubble}}{1926}]{Hubble26}
{Hubble} E.~P.,  1926, \apj, 64, 321

\bibitem[\protect\citeauthoryear{{Huchra}}{{Huchra}}{1977}]{Huchra77}
{Huchra} J.~P.,  1977, \apjs, 35, 171

\bibitem[\protect\citeauthoryear{{Koleva}, {Prugniel} \& {De Rijcke}}{{Koleva}
  et~al.}{2008}]{KolevaEtAl08}
{Koleva} M.,  {Prugniel} P.,    {De Rijcke} S.,  2008, Astronomische
  Nachrichten, 329, 968

\bibitem[\protect\citeauthoryear{{Komatsu} et~al.,}{{Komatsu}
  et~al.}{2011}]{KomatsuEtAl11}
{Komatsu} E.,  et~al., 2011, \apjs, 192, 18

\bibitem[\protect\citeauthoryear{{Kroupa}}{{Kroupa}}{2007}]{Kroupa07}
{Kroupa} P.,  2007, ArXiv Astrophysics e-prints astro-ph/0703124

\bibitem[\protect\citeauthoryear{{Lan{\c c}on} \& {Wood}}{{Lan{\c c}on} \&
  {Wood}}{2000}]{LanconWood00}
{Lan{\c c}on} A.,  {Wood} P.~R.,  2000, \aaps, 146, 217

\bibitem[\protect\citeauthoryear{{Le Borgne} et~al.,}{{Le Borgne}
  et~al.}{2003}]{LeBorgneEtAl03}
{Le Borgne} J.-F.,  et~al., 2003, \aaps, 402, 433

\bibitem[\protect\citeauthoryear{{Lejeune}, {Cuisinier} \& {Buser}}{{Lejeune}
  et~al.}{1997}]{LejeuneEtAl97}
{Lejeune} T.,  {Cuisinier} F.,    {Buser} R.,  1997, \aaps, 125, 229

\bibitem[\protect\citeauthoryear{{Lejeune}, {Cuisinier} \& {Buser}}{{Lejeune}
  et~al.}{1998}]{LejeuneEtAl98}
{Lejeune} T.,  {Cuisinier} F.,    {Buser} R.,  1998, \aaps, 130, 65

\bibitem[\protect\citeauthoryear{{Lintott} et~al.,}{{Lintott}
  et~al.}{2011}]{LintottEtAl11}
{Lintott} C.,  et~al., 2011, \mnras, 410, 166

\bibitem[\protect\citeauthoryear{{Lintott} et~al.,}{{Lintott}
  et~al.}{2008}]{LintottEtAl08}
{Lintott} C.~J.,  et~al., 2008, \mnras, 389, 1179

\bibitem[\protect\citeauthoryear{{MacArthur}, {McDonald}, {Courteau} \&
  {Jes{\'u}s Gonz{\'a}lez}}{{MacArthur} et~al.}{2010}]{MacArthurEtAl10}
{MacArthur} L.~A.,  {McDonald} M.,  {Courteau} S.,    {Jes{\'u}s Gonz{\'a}lez}
  J.,  2010, \apj, 718, 768

\bibitem[\protect\citeauthoryear{{Maller}, {Berlind}, {Blanton} \&
  {Hogg}}{{Maller} et~al.}{2009}]{MallerEtAl09}
{Maller} A.~H.,  {Berlind} A.~A.,  {Blanton} M.~R.,    {Hogg} D.~W.,  2009,
  \apj, 691, 394

\bibitem[\protect\citeauthoryear{{Maraston}}{{Maraston}}{2005}]{Maraston05}
{Maraston} C.,  2005, \mnras, 362, 799

\bibitem[\protect\citeauthoryear{{Masters} et~al.,}{{Masters}
  et~al.}{2010a}]{MastersEtAl10a}
{Masters} K.~L.,  et~al., 2010a, \mnras, 404, 792

\bibitem[\protect\citeauthoryear{{Masters} et~al.,}{{Masters}
  et~al.}{2010b}]{MastersEtAl10}
{Masters} K.~L.,  et~al., 2010b, \mnras, 405, 783

\bibitem[\protect\citeauthoryear{{Mignoli} et~al.,}{{Mignoli}
  et~al.}{2009}]{MignoliEtAl09}
{Mignoli} M.,  et~al., 2009, \aap, 493, 39

\bibitem[\protect\citeauthoryear{{Oesch} et~al.,}{{Oesch}
  et~al.}{2010}]{OeschEtAl10}
{Oesch} P.~A.,  et~al., 2010, \apjl, 714, L47

\bibitem[\protect\citeauthoryear{{Panter}, {Heavens} \& {Jimenez}}{{Panter}
  et~al.}{2003}]{PanterEtAl03}
{Panter} B.,  {Heavens} A.~F.,    {Jimenez} R.,  2003, \mnras, 343, 1145

\bibitem[\protect\citeauthoryear{{Petrosian}}{{Petrosian}}{1976}]{Petrosian76}
{Petrosian} V.,  1976, \apjl, 209, L1

\bibitem[\protect\citeauthoryear{{Robaina}, {Hoyle}, {Gallazzi}, {Jimenez},
  {van der Wel} \& {Verde}}{{Robaina} et~al.}{2011}]{RobainaEtAl12}
{Robaina} A.~R.,  {Hoyle} B.,  {Gallazzi} A.,  {Jimenez} R.,  {van der Wel} A.,
     {Verde} L.,  2011, ArXiv e-prints

\bibitem[\protect\citeauthoryear{{Schawinski} et~al.,}{{Schawinski}
  et~al.}{2007}]{SchawinskiEtAl07a}
{Schawinski} K.,  et~al., 2007, \apjs, 173, 512

\bibitem[\protect\citeauthoryear{{Schawinski} et~al.,}{{Schawinski}
  et~al.}{2009}]{SchawinskiEtAl09b}
{Schawinski} K.,  et~al., 2009, \mnras, 396, 818

\bibitem[\protect\citeauthoryear{{Schawinski}, {Lintott}, {Thomas}, {Sarzi},
  {Andreescu}, {Bamford}, {Kaviraj}, {Khochfar}, {Land}, {Murray}, {Nichol},
  {Raddick}, {Slosar}, {Szalay}, {Vandenberg} \& {Yi}}{{Schawinski}
  et~al.}{2009}]{SchawinskiEtAl09}
{Schawinski} K.,  {Lintott} C.,  {Thomas} D.,  {Sarzi} M.,  {Andreescu} D.,
  {Bamford} S.~P.,  {Kaviraj} S.,  {Khochfar} S.,  {Land} K.,  {Murray} P.,
  {Nichol} R.~C.,  {Raddick} M.~J.,  {Slosar} A.,  {Szalay} A.,  {Vandenberg}
  J.,    {Yi} S.~K.,  2009, \mnras, 396, 818

\bibitem[\protect\citeauthoryear{{Schlegel}, {Finkbeiner} \&
  {Davis}}{{Schlegel} et~al.}{1998}]{SchlegelEtAl98}
{Schlegel} D.~J.,  {Finkbeiner} D.~P.,    {Davis} M.,  1998, \apj, 500, 525

\bibitem[\protect\citeauthoryear{{Skelton}, {Bell} \& {Somerville}}{{Skelton}
  et~al.}{2012}]{SkeltonEtAl12}
{Skelton} R.~E.,  {Bell} E.~F.,    {Somerville} R.~S.,  2012, \apj, 753, 44

\bibitem[\protect\citeauthoryear{{Skibba} et~al.,}{{Skibba}
  et~al.}{2009}]{SkibbaEtAl09c}
{Skibba} R.~A.,  et~al., 2009, \mnras, 399, 966

\bibitem[\protect\citeauthoryear{{Stoughton} et~al.,}{{Stoughton}
  et~al.}{2002}]{StoughtonEtAl02}
{Stoughton} C.,  et~al., 2002, \aj, 123, 485

\bibitem[\protect\citeauthoryear{{Strateva} et~al.,}{{Strateva}
  et~al.}{2001}]{StratevaEtAl01}
{Strateva} I.,  et~al., 2001, \aj, 122, 1861

\bibitem[\protect\citeauthoryear{{Strauss} et~al.,}{{Strauss}
  et~al.}{2002}]{StraussEtAl02}
{Strauss} M.~A.,  et~al., 2002, \aj, 124, 1810

\bibitem[\protect\citeauthoryear{{Thomas} \& {Davies}}{{Thomas} \&
  {Davies}}{2006}]{ThomasEtAl06}
{Thomas} D.,  {Davies} R.~L.,  2006, \mnras, 366, 510

\bibitem[\protect\citeauthoryear{{Thomas}, {Maraston}, {Schawinski}, {Sarzi} \&
  {Silk}}{{Thomas} et~al.}{2010}]{ThomasEtAl10}
{Thomas} D.,  {Maraston} C.,  {Schawinski} K.,  {Sarzi} M.,    {Silk} J.,
  2010, \mnras, 404, 1775

\bibitem[\protect\citeauthoryear{{Tojeiro}, {Heavens}, {Jimenez} \&
  {Panter}}{{Tojeiro} et~al.}{2007}]{TojeiroEtAl07}
{Tojeiro} R.,  {Heavens} A.~F.,  {Jimenez} R.,    {Panter} B.,  2007, \mnras,
  381, 1252

\bibitem[\protect\citeauthoryear{{Tojeiro} \& {Percival}}{{Tojeiro} \&
  {Percival}}{2011}]{TojeiroEtAl11b}
{Tojeiro} R.,  {Percival} W.~J.,  2011, \mnras, 417, 1114

\bibitem[\protect\citeauthoryear{{Tojeiro}, {Percival}, {Heavens} \&
  {Jimenez}}{{Tojeiro} et~al.}{2011}]{TojeiroEtAl11}
{Tojeiro} R.,  {Percival} W.~J.,  {Heavens} A.~F.,    {Jimenez} R.,  2011,
  \mnras, 413, 434

\bibitem[\protect\citeauthoryear{{Tojeiro}, {Wilkins}, {Heavens}, {Panter} \&
  {Jimenez}}{{Tojeiro} et~al.}{2009}]{TojeiroEtAl09}
{Tojeiro} R.,  {Wilkins} S.,  {Heavens} A.~F.,  {Panter} B.,    {Jimenez} R.,
  2009, \apjs, 185, 1

\bibitem[\protect\citeauthoryear{{Wolf} et~al.,}{{Wolf}
  et~al.}{2009}]{WolfEtAl09}
{Wolf} C.,  et~al., 2009, \mnras, 393, 1302

\bibitem[\protect\citeauthoryear{{York} et~al.,}{{York}
  et~al.}{2000}]{YorkEtAl00}
{York} D.~G.,  et~al., 2000, \aj, 120, 1579

\end{thebibliography}

\appendix

\section{Results obtained with BC03 stellar population models}\label{sec:appendix_BC03}

\begin{figure}
\begin{center}
\includegraphics[width=2.5in,angle=90]{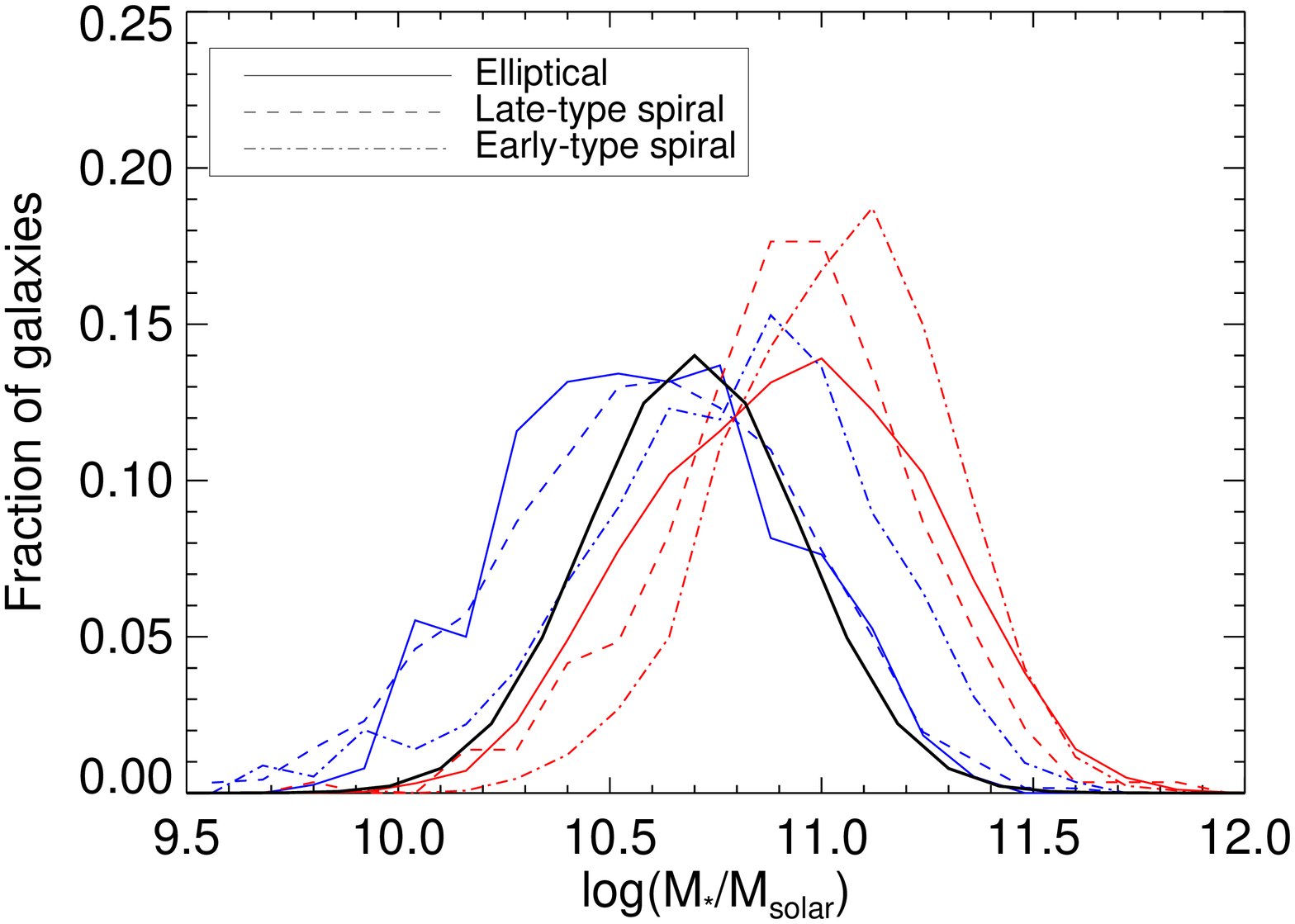}
\caption{The normalised distribution of stellar masses for blue elliptical (blue solid), blue spiral (blue dashed), red elliptical (red solid) and red spiral (red dashed) galaxies as per the database of \protect\cite{TojeiroEtAl09}, computed using the SSP models of BC03. This Figure shows clearly how stellar mass follows colour, rather than morphology. The black curve shows the stellar mass distribution to which all samples are weighted, peaking and centred at  $M_* \sim 10^{10.7} M_\odot$; this allows us to eliminate stellar mass as a driver of physical differences we observe between the four samples of galaxies.} 
\label{fig:stellar_masses_BC03}
\end{center}
\end{figure}

\begin{figure*}
\begin{center}
\includegraphics[width=7in]{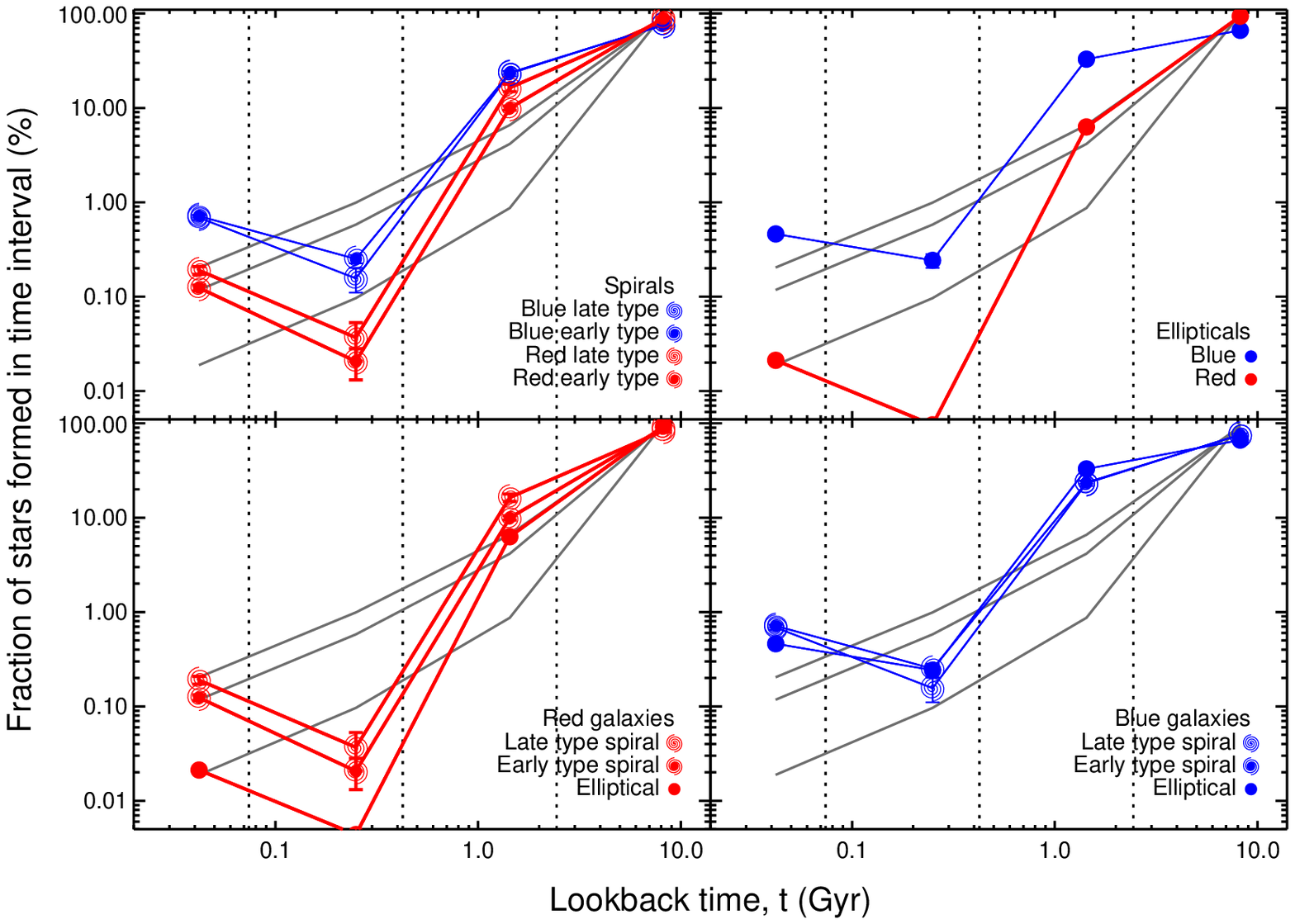}
\caption{Same as Fig.~\ref{fig:SFF_M05}, but computed using the BC03 SSP models. The striking dip seen at ages of roughly 1Gyr in the average star-formation fractions is a common feature of these models, and seen also in \protect\cite{PanterEtAl03,TojeiroEtAl09}.} 
\label{fig:SFF_BC03}
\end{center}
\end{figure*}

\begin{figure}
\begin{center}
\includegraphics[width=6in]{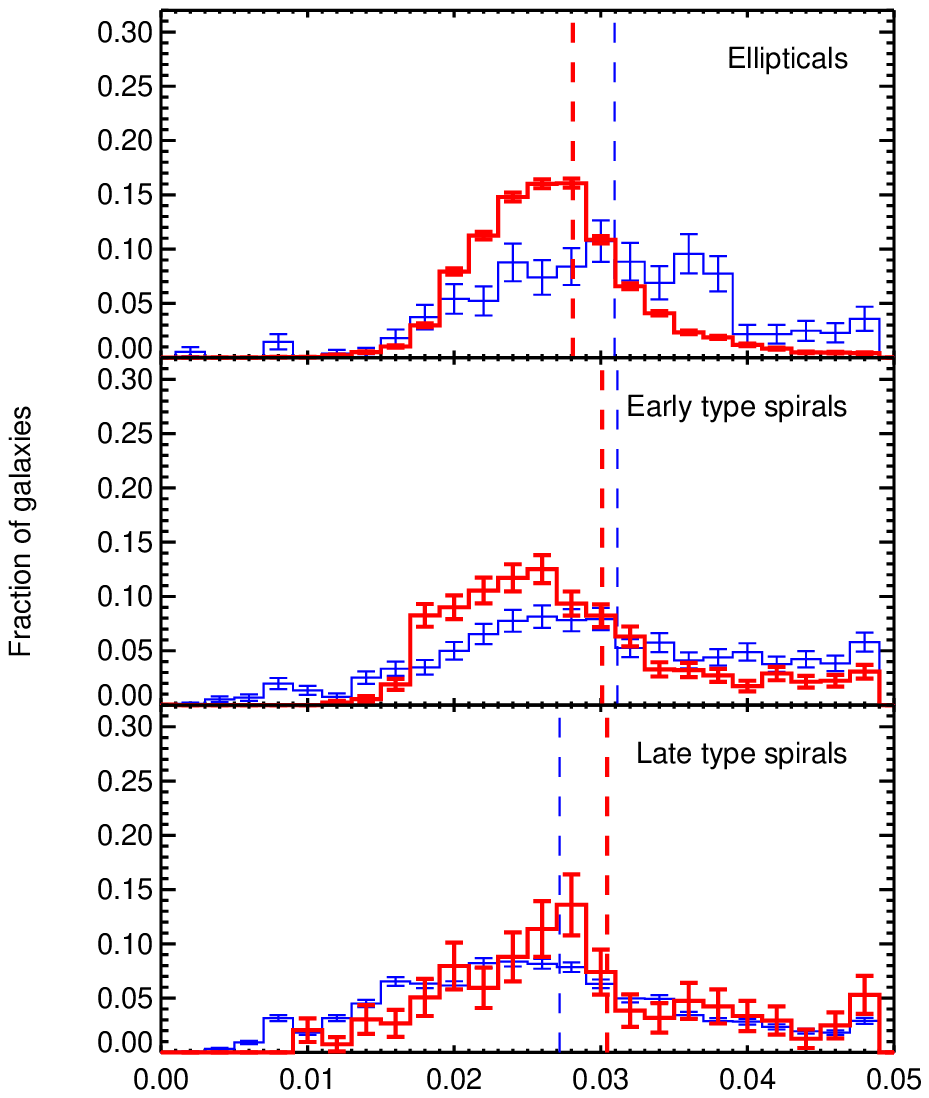}
\caption{The distribution of the mass-weighted stellar metallicity for red and blue galaxies (thicker and thinner lines, respectively) for all three morphological samples, using BC03 modelling. The vertical dashed line shows the mean value of each distribution. The error bars are Poisson errors. The scatter is visibly larger in the case of BC03 models, and the mass-weighted metallicities are on average larger. This trend was already seen in \protect\cite{TojeiroEtAl09,TojeiroEtAl11}, especially at solar metallicities and above. } 
\label{fig:metallicity_BC03}
\end{center}
\end{figure}

\begin{figure*}
\begin{center}
\includegraphics[angle=90,width=5.5in]{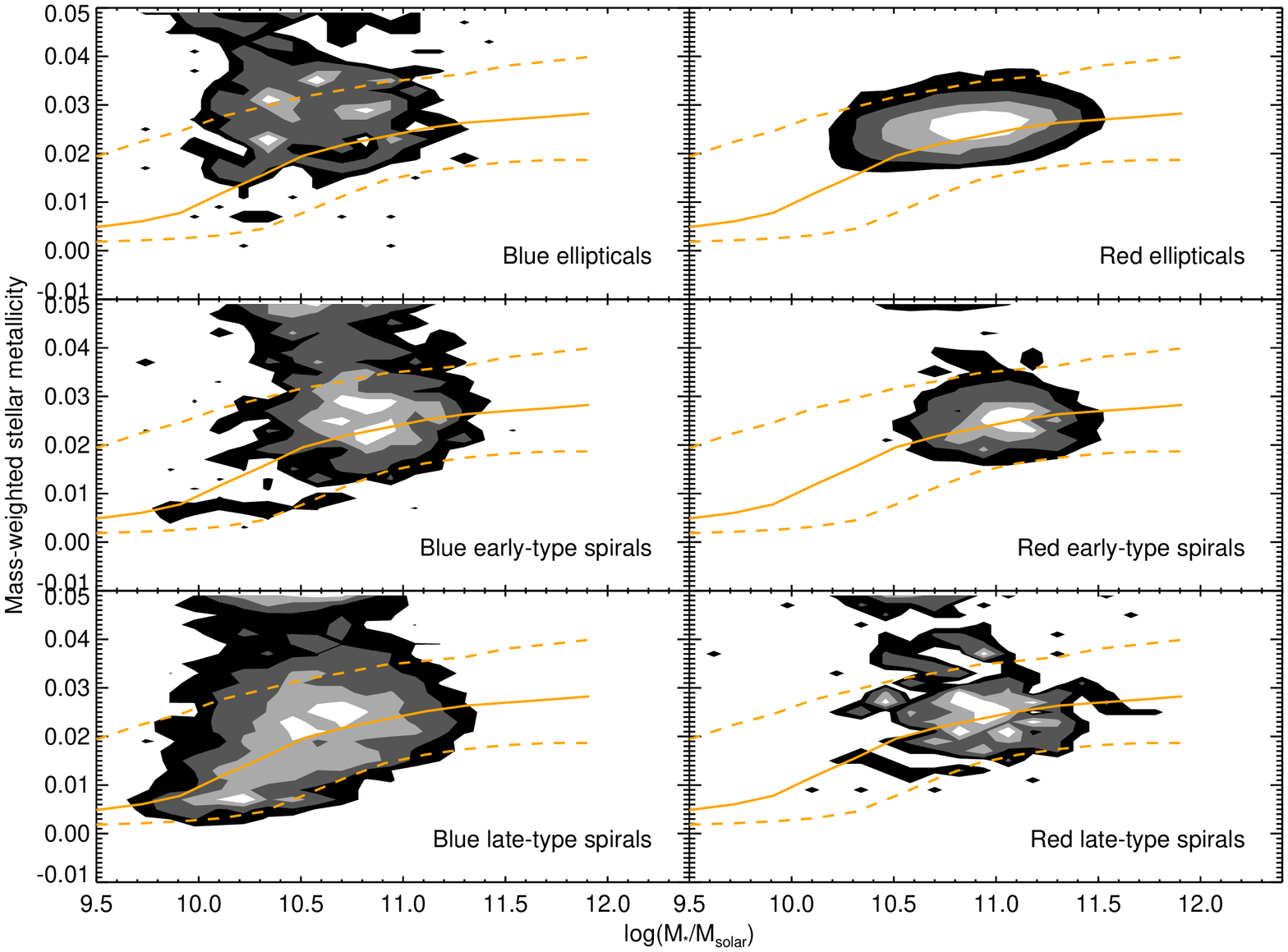}
\caption{Stellar mass vs metallicity relation for the four samples. Left hand side shows blue galaxies, and right hand since red galaxies. The orange solid and dashed lines show the stellar mass - metallicity relation measured by \protect\cite{GallazziEtAl05}. The mass-metallicity relation is tighter in the case of the M05 models for all galaxy samples. } 
\label{fig:2D_metallicity_BC03}
\end{center}
\end{figure*}

\begin{figure}
\begin{center}
\includegraphics[width=6in]{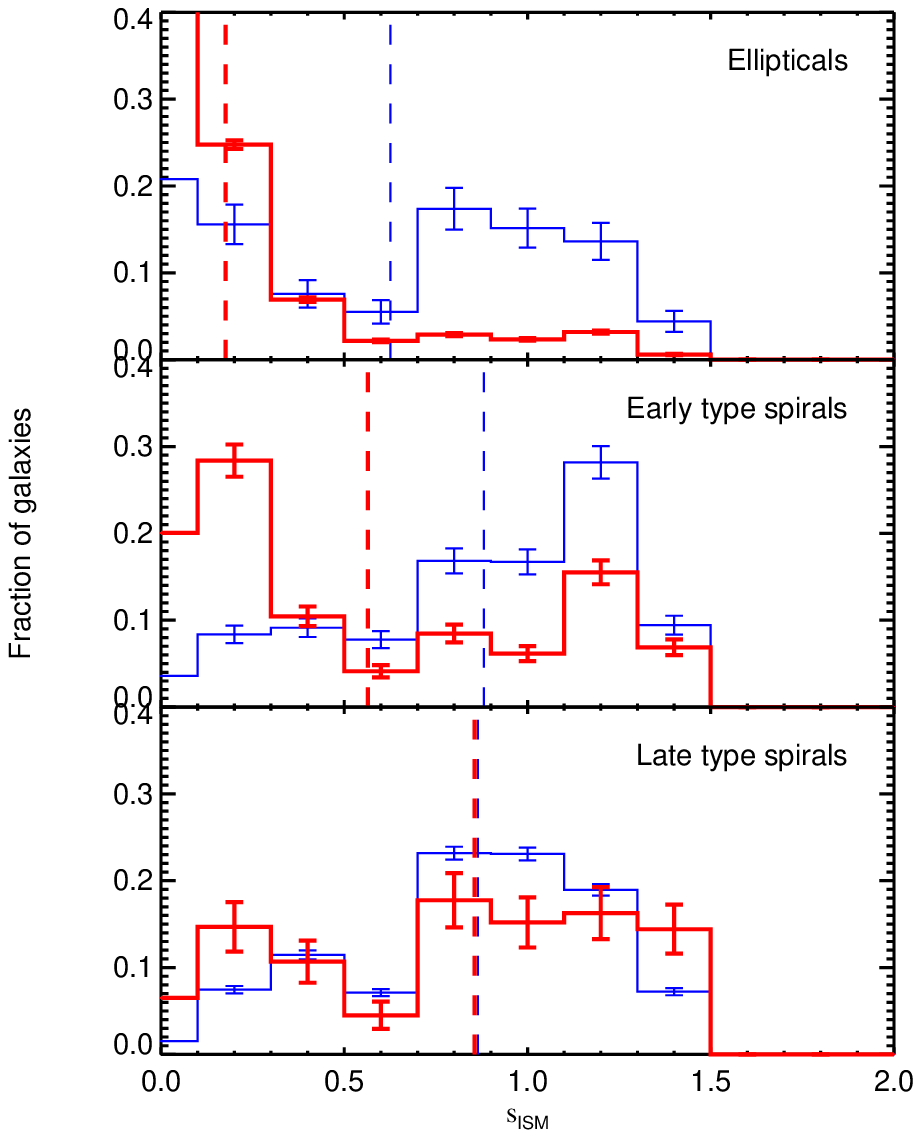}
\caption{The distribution of the optical depth $\tau_{ISM}$ for red and blue galaxies (thicker and thinner lines, respectively) for all three morphological samples, using BC03 modelling. The vertical dashed line shows the mean value of each distribution. The error bars are Poisson errors. } 
\label{fig:dust_BC03}
\end{center}
\end{figure}

\end{document}